\documentclass[twocolumn,twoside]{IEEEtran}
\usepackage[utf8]{inputenc} % The pack­age trans­lates var­i­ous stan­dard and other in­put en­cod­ings into a `LaTeX in­ter­nal lan­guage'
\usepackage[T1]{fontenc} % al­lows the user to se­lect font en­cod­ings, and for each en­cod­ing pro­vides an in­ter­face to `font-en­cod­ing-spe­cific' com­mands for each font
\usepackage[cmex10]{amsmath} % Use the [cmex10] option to ensure complicance
\usepackage{amssymb} % provides an extended symbol collection
\usepackage{mathtools} % optional corrections to amsmath, but recommended
\usepackage{amsfonts} % ex­tended set of fonts for use in math­e­mat­ics

\usepackage{mleftright}\mleftright % defines variants \mleft and \mright of \left and \right, that make the delimiters act as \mathopen and \mathclose. These commands address spacing difficulties in subformulas.

\usepackage[dvips]{graphicx} % for including graphics
\graphicspath{{./}}
\DeclareGraphicsExtensions{.eps}

\usepackage[dvipsnames]{xcolor} % provides easy driver-independent access to several kinds of colors, tints, shades, tones, and mixes of arbitrary colors; options: [xetex]

\usepackage{bbm} % double-striked font for most symbols
\usepackage{xfrac} % diagonal fractions using \sfrac
\usepackage{cancel} % crossed-over equations; cancel-to-zero, etc. Options: [makeroom]
\usepackage{wrapfig} % typesetting a narrow float at the edge of the text, and making the text wrap around it.
\usepackage{tensor} % Al­lows the user to set ten­sor-style su­per- and sub­scripts with off­sets be­tween suc­ces­sive in­dices

\usepackage{theorem}
\usepackage{cite}

\usepackage{url} % The com­mand \url is a form of ver­ba­tim com­mand that al­lows line­breaks at cer­tain char­ac­ters or com­bi­na­tions of char­ac­ters, ac­cepts re­con­fig­u­ra­tion, and can usu­ally be used in the ar­gu­ment to an­other com­mand

\usepackage{psfrag} % Al­lows LaTeX construc­tions (equa­tions, pic­ture environments, etc.) to be pre­cisely superim­posed over En­cap­su­lated PostScript fig­ures

\usepackage[inline]{enumitem} % list customization; use [inline] to enable horizontal lists with *-ed environments

\usepackage{array} % extended  implementation  of  the  LaTeX array– and tabular–environments

\newcolumntype{P}[1]{>{\centering\arraybackslash}p{#1}}

\usepackage{aliascnt} % allows for a single counter in \autoref'ed theorems

\usepackage[hidelinks,linktocpage]{hyperref}

\usepackage{tikz}
\usetikzlibrary{shapes.geometric, arrows}

\usepackage{xparse} % used for conditional macro definitions

\usepackage{ifthen} % Ifthen is a sep­a­rate pack­age within the LaTeX dis­tri­bu­tion; while it will al­ways be present in a LaTeX dis­tri­bu­tion, a \usep­a­ck­age com­mand is al­ways needed to load it

\newcommand{\mynewtheorem}[2]{
  \newaliascnt{#1}{dummy}
  \newtheorem{#1}[#1]{#2}
  \aliascntresetthe{#1}
  \expandafter\def\csname #1autorefname\endcsname{#2}
}

\theoremstyle{plain}
\theorembodyfont{\normalfont}

  \mynewtheorem{theorem}{Theorem$\!$}
  \mynewtheorem{lemma}{Lemma$\!$}
  \mynewtheorem{corollary}{Corollary$\!$}

  \mynewtheorem{definition}{Definition$\!$}

  \mynewtheorem{xmpl}{Example$\!$}
  \newenvironment{example}{\begin{xmpl}\hspace*{-1ex}}{\hfill$\Box$\end{xmpl}}

\newtheorem{cnstr}{Construction$\!$}

\newenvironment{construction}{\begin{cnstr}}{\end{cnstr}} % removed the QED because it was acting strange

\newtheorem{myalgo}{Algorithm$\!$}

\makeatletter
\renewcommand{\@endtheorem}{\endtrivlist}
\makeatother

\makeatletter
\renewcommand{\thefigure}{{\@arabic\c@figure}}
\renewcommand{\fnum@figure}{{\bf Figure\,\thefigure}}
\makeatother

\usepackage{makecell}   %%% What does it do?

\renewcommand{\le}{\leqslant}
\renewcommand{\leq}{\leqslant}
\renewcommand{\ge}{\geqslant}
\renewcommand{\geq}{\geqslant}

\newcommand{\cD}{\mathcal{D}}

\newcommand{\cW}{\mathcal{W}}

\renewcommand{\Bbb}{\mathbb}

\newcommand{\N}{{\Bbb N}}

\newcommand{\Z}{{\Bbb Z}}

\newcommand{\F}{{\Bbb F}}

\DeclarePairedDelimiter\abs{\lvert}{\rvert}
\DeclarePairedDelimiter\ceilenv{\lceil}{\rceil}
\DeclarePairedDelimiter\floorenv{\lfloor}{\rfloor}
\DeclarePairedDelimiter\parenv{\lparen}{\rparen}
\DeclarePairedDelimiter\sparenv{\lbrack}{\rbrack}
\DeclarePairedDelimiter\bracenv{\lbrace}{\rbrace}
\DeclarePairedDelimiterX\mathset[2]{\lbrace}{\rbrace}{#1 \mathrel{}\delimsize\vert\mathrel{} #2}
\DeclarePairedDelimiterX\inner[2]{\langle}{\rangle}{#1 \mathrel{},\mathrel{} #2}

\DeclareDocumentCommand\norm{ o m }{
    \IfNoValueTF{#1}
        {\left\Vert#2\right\Vert}
        {\left\Vert#2\right\Vert_{#1}}
}

\DeclareDocumentCommand\der{ o m o }{
    \IfNoValueTF{#1}
        {
            \IfNoValueTF{#3}
                {\frac{d}{d{#2}}}
                {\frac{d{#3}}{d{#2}}}
        }
        {\parenv*{\frac{d}{d{#2}}}^{#1}\IfNoValueTF{#3}{}{#3}}
}
\DeclareDocumentCommand\partder{ o m m }{
    \IfNoValueTF{#1}
        {\frac{\partial{#3}}{\partial{#2}}}
        {\frac{\partial^{#1}{#3}}{{\partial{#2}}^{#1}}}
}
\DeclareDocumentCommand\df{ o m o }{
    d\IfNoValueTF{#1}{}{^{#1}}{#2}\IfNoValueTF{#3}{}{_{#3}}
}

\newcommand{\eqdef}{\triangleq}
\newcommand{\deq}{\triangleq}

\DeclareMathOperator{\irr}{Irr}
\DeclareMathOperator{\rll}{RLL}
\DeclareMathOperator{\wt}{wt}

\DeclareMathOperator{\rt}{drt}
\newcommand{\ond}{\mathrm{1ND}}
\newcommand{\aux}{\mathrm{aux}}
\newcommand{\dc}{D}

\begin{document}

%%%%%%%%%%%%%%%%%%%%%%%%%%%%%%%%%%%%%%%%%%%%%%%%%%%%%%%%%%%%%%%%%%%%%%
%%%%%%%%%%%%%%%%%%%%%%%%%%%%%%%%%%%%%%%%%%%%%%%%%%%%%%%%%%%%%%%%%%%%%%
\title{Single-Error Detection and Correction for Duplication and Substitution Channels}

%\author{
 % \IEEEauthorblockN{ Yuanyuan Tang\IEEEauthorrefmark{1}, 
  %                   Yonatan Yehezkeally\IEEEauthorrefmark{2}, 
 %                    Moshe Schwartz\IEEEauthorrefmark{3}, 
  %                   and Farzad Farnoud (Hassanzadeh)\IEEEauthorrefmark{4}} %\\
  %\IEEEauthorblockA{\IEEEauthorrefmark{1}%
   %                 Electrical \& Computer Engineering, 
    %                University of Virginia, 
     %               \texttt{yt5tz@virginia.edu}} %\\
  %\IEEEauthorblockA{\IEEEauthorrefmark{2}%
   %                 Electrical \& Computer Engineering, 
    %                Ben-Gurion University of the Negev,
     %               \texttt{yonatany@post.bgu.ac.il}} %\\
  %\IEEEauthorblockA{\IEEEauthorrefmark{3}%
   %                 Electrical \& Computer Engineering, 
    %                Ben-Gurion University of the Negev,
     %               \texttt{schwartz@ee.bgu.ac.il}} %\\
 % \IEEEauthorblockA{\IEEEauthorrefmark{4}%
 %                   Electrical \& Computer Engineering, 
  %                  University of Virginia, 
   %                 \texttt{farzad@virginia.edu}}
 % \thanks{This work was supported in part by NSF grants under grant %nos.~1816409 and~1755773, and a BSF grant under grant %no.~2017652.}\vspace{-2.0em}
%}

\author{Yuanyuan~Tang, Yonatan~Yehezkeally,~\IEEEmembership{Student~Member,~IEEE}, \\ Moshe~Schwartz,~\IEEEmembership{Senior Member,~IEEE,} and~Farzad~Farnoud,~\IEEEmembership{Member,~IEEE}%
\thanks{This paper was presented in part at ISIT 2019.}%
\thanks{Yuanyuan Tang is with the Department
    of Electrical and Computer Engineering, 
    University of Virginia, Charlottesville, VA, 22903, USA, (email: yt5tz@virginia.edu).}%
  \thanks{Yonatan Yehezkeally is with the School
    of Electrical and Computer Engineering, Ben-Gurion University of the Negev,
    Beer Sheva 8410501, Israel
    (e-mail: yonatany@bgu.ac.il).}%
  \thanks{Moshe Schwartz is with the School
    of Electrical and Computer Engineering, Ben-Gurion University of the Negev,
    Beer Sheva 8410501, Israel
    (e-mail: schwartz@ee.bgu.ac.il).}%
  \thanks{Farzad Farnoud (Hassanzadeh) is with the Department of Electrical and Computer Engineering and the Department of Computer Science, University of Virginia, Charlottesville, VA, 22903, USA, (email: farzad@virginia.edu).}% 
  \thanks{This work was supported in part by National Science Foundation (NSF) grants under grant nos.~1816409 and~1755773, and a U.S-Israel Binational Science Foundation (BSF) grant under grant no.~2017652.}%
  \thanks{Copyright (c) 2020 IEEE. Personal use of this material is permitted. However, permission to use this material for any other purposes must be obtained from the IEEE by sending a request to pubs-permissions@ieee.org.}
  }

\maketitle

\begin{abstract}
Motivated by mutation processes occurring in in-vivo DNA-storage 
applications, a channel that mutates stored strings by duplicating 
substrings as well as substituting symbols is studied. Two models of 
such a channel are considered: one in which the substitutions occur 
only within the duplicated substrings, and one in which the location 
of substitutions is unrestricted. Both error-detecting and 
error-correcting codes are constructed, which can handle correctly any 
number of tandem duplications of a fixed length $k$, and at most a 
single substitution occurring at any time during the mutation process.
\end{abstract}


\section{Introduction}

\IEEEPARstart{R}{ecent} advances in DNA sequencing and synthesis 
technologies have increased the potential of DNA as a data-storage 
medium. In addition to its high data density, data storage in DNA 
provides a long-lasting alternative to current storage media. 
Furthermore, given the need for accessing biological data stored in 
DNA of living organisms, technologies for retrieving data from DNA 
will not become obsolete, unlike flash memory, magnetic disks, and 
optical disks.

Data can be stored in DNA \emph{in vitro} or \emph{in vivo}. While the 
former will likely provide a higher density, the latter can provide a 
more reliable and cost-effective replication method, as well as a 
protective shell~\cite{shipman2017}. In-vivo storage also has 
applications such as watermarking genetically modified organisms. This 
technology was recently demonstrated experimentally using CRISPR/Cas 
gene editing~\cite{shipman2016,shipman2017}. One of the challenges of 
this technology is that a diverse set of errors are possible, 
including substitutions, duplications, insertions, and deletions. 
Duplication errors, in particular, have been previously studied by a 
number of recent works, including~\cite{jain2017e, LenWacYaa19, 
sala2017exact, mahdavifar2017, kovacevic2018}, among others. This 
paper focuses on error-control codes for duplication and substitution 
errors. 

In a (tandem) duplication event, a substring of the DNA sequence, the 
\emph{template}, is duplicated and the resulting \emph{copy} is 
inserted into the sequence next to the template~\cite{zhou2014}. 
Evidence of this process is found in the genomes of many organisms as 
patterns that are repeated multiple times~\cite{farnoud2018}. In a 
substitution event, a symbol in the sequence is changed to another 
symbol of the alphabet. It has been observed that point mutations such 
as substitutions are more common in tandem repeat regions of the 
genomes~\cite{pumpernik2008}. We consider two models for combined 
duplication and substitution errors. In the first model, called the 
\emph{noisy-duplication model}, the copy is a noisy version of the 
template. Noisy duplications in this model can be viewed as exact 
duplications followed by substitutions that are restricted to the 
newly added copy. Hence, this model is also referred to as the 
\emph{restricted-substitution model}. We also consider an 
\emph{unrestricted-substitution model}, which relaxes the noisy 
duplication model by allowing substitutions at any position in the 
sequence.

In this paper we construct both error-detecting and error-correcting 
codes, which are capable of correctly handling any number of tandem 
duplications of a fixed length~$k$, and at most a single substitution 
error, which occurs at any stage during the sequence of duplication 
events.
The main approach in both cases is to reverse the duplication process 
while accounting for the single substitution (which may spuriously 
create the appearance of a duplication that never happened, or 
eliminate one that did). Different challenges are also presented by 
the possible locations for substitutions. We bring these differences 
to light by providing a construction for an error-detecting code for 
the restricted substitution model, and an error-correcting code for 
the unrestricted substitution model.

Our main contributions are the following:
\begin{itemize}
\item
We present an upper bound on the minimum required redundancy cost for 
detecting a single restricted substitution, over the necessary rate 
loss required to correct an unlimited number of duplication events, in 
\autoref{thm:1SDbounds}. That cost is upper bounded by $O(\log(n-k))$.

\item
Through \autoref{Const:code} and \autoref{Const:codeconstructive}, we 
also show that the redundancy cost (over the rate loss due to 
duplication noise) is upper bounded by $O(\log(k))$ and $O(k)$, 
respectively. While the former guarantees larger codes, it is 
nonconstructive, as opposed to the latter. In the likely regime where 
$k$ is fixed, both require only $O(1)$ extra redundancy.

\item
Through \autoref{Const:1sdetcode}, we show that the redundancy cost of 
detecting a single unrestricted substitution (again, over the rate 
loss due to duplication noise) is upper bounded by 
$O\parenv*{\log(k^2 n)}$.

\item
Finally, in \autoref{cor:tan-sing_mut} and \autoref{thm:ecc}, we 
correct a single unrestricted substitution in addition to any number 
of duplications, but we incur further rate loss.

\end{itemize}

This paper is organized as follows. In Section~\ref{sec:prelim}, we 
provide the notation as well as relevant background and known results. 
In Section \ref{sec:detect}, we construct error-detecting codes for 
the restricted substitution model. 
In Section~\ref{sec:detect_un_subst}, we introduce error-detecting 
codes for unrestricted substitution channels. 
Finally, in Section~\ref{sec:correct}, we give a construction for an 
error-correcting code for the unrestricted substitution model. 
We conclude with a discussion of the results, and point out some open 
problems, in \autoref{sec:conclusion}.

\section{Notation and Preliminaries}
\label{sec:prelim}

Throughout the paper, we assume that the alphabet $\Sigma$ is a unital 
ring of size $q\geq 2$ (e.g., $\Z_q$ or, when $q$ is a prime power, 
$\F_q$). Thus, addition (or subtraction) and multiplications of 
letters from the alphabet are well-defined. The set of finite strings 
and strings of length at least $k$ over $\Sigma$ is denoted $\Sigma^*$ 
and $\Sigma^{\geq k}$, respectively. The concatenation of two strings, 
$u,v\in\Sigma^*$ is denoted by $uv$, and $u^k$ denotes concatenating 
$k$ copies of $u$. To avoid confusion, the multiplication in the ring 
is denoted as $a\cdot b$. We say $y\in\Sigma^*$ is a substring of 
$w\in\Sigma^*$ if there exist $x,z\in\Sigma^*$ such that $w=xyz$. 

The length (number of letters) of $u$ is denoted by $\abs{u}$, and for 
$a\in\Sigma$, we use $\abs{u}_a$ to denote the number of occurrences 
of $a$ in $u$. The Hamming weight of $u$ is denoted by $\wt(u)$, and 
if $\abs{u}=\abs{v}$ we use $d(u,v)$ to denote the Hamming distance 
between $u$ and $v$. If the need arises to refer to specific positions 
in words, positions are numbered $1,2,\dots$.

A (tandem) \emph{duplication} of length $k$ duplicates a substring of 
length $k$ and inserts it in tandem into the string, namely, the copy 
immediately follows the template. For example, from $uvw$, where 
$\abs*{v}=k$, we may obtain $uvvw$.  As an example for $k=3$ and 
alphabet $\Sigma=\Z_3$, consider 
\begin{equation}\label{eq:ex0}
    \begin{split}
        x   = 1012121&\to x'  = 1012\underline{012}121,
    \end{split}
\end{equation}
where the underlined part is the copy. Since throughout the paper all 
duplications considered will be in tandem and of length~$k$, we shall 
just use the term ``duplication'' to avoid cumbersome terminology.

The analysis of duplication errors will be facilitated by the 
$k$-\emph{discrete-derivative} transform, defined in 
\cite{hassanzadeh2016capacity} in the following way. For
 $x\in \Sigma^{\geq k}$, we define 
 $\phi(x) \deq \hat\phi(x)\bar\phi(x)$, where 
\begin{align*}
    \hat\phi(x)&\deq x_1\dotsm x_k,&
    \bar\phi(x)&\deq x_{k+1}\dotsm x_n - x_1\dotsm x_{n-k},
\end{align*}
in which subtraction is performed entry-wise over $\Sigma$.
We note that $\phi(\cdot)$ is a bijection. 
The duplication length $k$ is implicit in the definition of $\phi$. 
For a set of strings $S$, we define 
$\phi(S)\eqdef\mathset{\phi(s)}{s\in S}$.

Let $x'$ be obtained through a tandem duplication of length $k$ from 
$x$. 
It is not difficult to see that $\hat\phi(x)=\hat\phi(x')$ and that 
$\bar\phi(x')$ can be obtained from $\bar\phi(x)$ by inserting $0^k$ 
in an appropriate position~\cite{jain2017e}. For the example given 
in~\eqref{eq:ex0}, 
 \begin{equation*}
    \begin{split}\label{eq:ex1}
        x   = 1012121&\to x'  = 1012\underline{012}121\\
        \phi(x) = 101,1112 &\to \phi(x') = 101,1\underline{000}112
    \end{split}
\end{equation*}
Here, a comma separates the two parts of $\phi$ for clarity.

Sometimes duplications are \emph{noisy} and the duplicated symbols are
different from the original symbols. (Unless otherwise stated 
duplications are assumed to be exact.) We only consider the case where 
a single symbol is different. We view a noisy duplication as a 
duplication followed by a substitution in the duplicated substring. 
Continuing the example, the duplication resulting in $x'$ may be 
followed by a substitution, 
\begin{equation*}
    \begin{split}\label{eq:ex2}
        x'  = 1012\underline{0}12121 &\to x'' = 1012\underline{1}12121,\\
        \phi(x')= 101,1\underline{0}00\underline{1}12 &\to \phi(x'')=101,1\underline{1}00\underline{0}12.
    \end{split}
\end{equation*}

We also consider unrestricted substitutions, which can occur at any 
position in the string, rather than only in a substring that is 
duplicated by the previous duplication. A substitution may be 
considered as the mapping $x\to x+a e_i$, where $e_i\in\Sigma^n$ is a 
standard unit vector at index $i$, and $a\in\Sigma$, $a\neq 0$. Since 
$\phi$ is linear over $\Sigma$ (i.e., $\phi(x + a e_i) = \phi(x) 
+ a \phi(e_i)$), we denote the transform of $e_i$ as $\epsilon_i\deq 
\phi(e_i)$, and observe that $\epsilon_i = e_i - e_{i+k}$ for 
$i\leq n-k$ and $\epsilon_i = e_i$ for $n-k<i\leq n$. We note that 
substitutions might affect two positions in the $\phi$-transform 
domain.

Let $\dc^{t(p)}(x)$ (for $t\ge p$) denote the set of strings that can 
be obtained from $x$ through $t$ tandem duplications, $p$ of which are 
noisy (in any order), with each noisy duplication containing a single 
substitution. $\dc^{t(p)}$ is called a \emph{descendant cone} of $x$.
Continuing our earlier examples, we have $x'\in \dc^{1(0)}(x)$ 
and $x''\in \dc^{1(1)}(x)$.
We further define 
\begin{equation}\label{eq:des-cone}
\begin{split}
    \dc^{*(p)}(x)\eqdef\bigcup_{t=p}^\infty \dc^{t(p)}(x), \quad
    \dc^{*(P)}(x)\eqdef\bigcup_{p\in P} \dc^{*(p)}(x),\\
\end{split}
\end{equation}
where $P$ is a subset of non-negative integers. We denote $P=\{0,1\}$ 
as $\le1$. 

We define $\dc^{t,p}(x)$ to be the set of strings obtained from $x$ 
through $t$ tandem duplications and $p$ substitutions, where 
substitutions can occur in any position (and so we do not require 
$t\ge p$), and at any stage during the duplication sequence. We extend 
this definition similarly to~\eqref{eq:des-cone}. Obviously, for all 
$x\in\Sigma^*$,
\[ \dc^{t(0)}(x)=\dc^{t,0}(x).\]

For a string $z\in\Sigma^*$, $\mu(z)$ is obtained by removing all 
copies of $0^k$ from $z$. Specifically, for
\begin{equation*}
       z=0^{m_{0}}w_{1}0^{m_{1}}w_{2}\cdots w_{d}0^{m_{d}},
\end{equation*}
where $m_{i}$ are non-negative integers and $w_{i}\in \Sigma\setminus
\{0\}$ are nonzero symbols, we define
\begin{equation*}%\label{eq:def_irr}
       \mu(z)\eqdef 0^{m_{0}\bmod k}w_{1}0^{m_{1} \bmod k}w_{2}\cdots w_{d}0^{m_{d}\bmod k},
\end{equation*}
where $k$ is implicit in the notation $\mu(z)$.
For example, if $z = 1000112 = \bar{\phi}(x')$ from our earlier 
example, with $k=3$, then $\mu(z) = 1112$; note, then, that in that 
example, $\mu(z) = \bar{\phi}(x)$.

Define the \emph{duplication root} $\rt(x)$ of $x$ as the unique 
string obtained from $x$ by removing all tandem repeats of length~$k$, 
where the dependence on $k$ is implicit in the notation. For proof of 
the uniqueness of $\rt(x)$ see, e.g.,~\cite{jain2017e}. Note that 
\[\phi(\rt(x))=\hat\phi(x)\mu(\bar\phi(x))\]
(see \cite{jain2017e}); indeed, in our running example, $x=\rt(x')$.
For a set of strings $S$, we define 
\[\rt(S)\eqdef\mathset{\rt(s)}{s\in S}.\]

A string $x$ is \emph{irreducible} if $x=\rt(x)$. The set of 
irreducible strings of length $n$ is denoted $\irr(n)$, where the 
duplication length $k$ is again implicit. We denote by $\rll(m)$ the 
set of strings in $\Sigma^m$ that do not contain $0^k$ as a substring, 
i.e., the $(0,k-1)$-run-length limited (RLL) constrained strings of 
length $m$. A string $x$ of length $n$ is irreducible if and only if 
$\bar\phi(x)\in\rll({n-k})$.

A code $C\subseteq\Sigma^n$ that can correct any number of 
$k$-duplication errors is called a $k$-\emph{duplication code}. We 
note that a code is a $k$-duplication code if and only if no two 
distinct codewords $c_1,c_2\in C$ have a common descendant, namely,
\begin{equation}\label{eq:single_root}
    \dc^{*,0}(c_1)\cap \dc^{*,0}(c_2)=\varnothing.
\end{equation}
It was proved in \cite{jain2017e} that this condition is equivalent to 
all codewords having distinct roots:
\begin{theorem}
\label{th:rootint}
(\hspace{1sp}\cite{jain2017e}) For all strings, $x_1,x_2\in\Sigma^*$,
\[ \dc^{*,0}(x_1)\cap \dc^{*,0}(x_2)\neq \varnothing\]
if and only if $\rt(x_1)=\rt(x_2)$.
\end{theorem}

Using Theorem~\ref{th:rootint}, it was suggested in~\cite{jain2017e} 
that error-correcting codes that protect against any number of 
duplications may be obtained simply by taking irreducible words as 
codewords. Up to a minor tweaking, this strategy was shown 
in~\cite{jain2017e} to produce optimal codes.

Finally, we define the \emph{redundancy} of a code 
$C\subseteq\Sigma^n$ as
\[
r(C) \deq n - \log_q\abs*{C} = n - \log_{\abs*{\Sigma}}\abs*{C},
\]
and the code's \emph{rate} as 
\[R(C)\deq 1 - \frac{r(C)}{n}.\]

\section{Restricted Error-Detecting Codes}
\label{sec:detect}

\subsection{The error model and the descendant cone}

In this section, we consider the case of noisy-duplication errors. Our 
goal is to correct errors consisting of any number of exact 
duplications, or detect the presence of a single noisy duplication, 
which contains only one substitution.
We refer to codes with this capability as \emph{1-noisy duplication 
(1ND)} detecting. Let us first be more precise in our definition:

\begin{definition}
A code $C\subseteq\Sigma^*$ is a 1ND-detecting code if there exists a 
decoding function $\cD:\Sigma^*\to C\cup\{\mathrm{error}\}$ such that 
if $c\in C$ was transmitted and $y\in\Sigma^*$ was received then 
$\cD(y)=c$ if only duplication errors occurred, and $\cD(y)\in
\{c,\mathrm{error}\}$ if exactly one of the duplication errors that 
occurred was noisy, where the noisy duplication could have occurred at 
any point in the sequence of the duplication errors.
\end{definition}

\begin{figure}
    \centering
    %{\includegraphics[trim=0in 5in 11in 0in,clip=on,width=.5\columnwidth]{1ND-def.pdf}}
    {\includegraphics[width=.5\columnwidth]{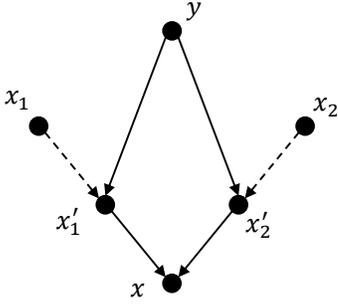}}
    \caption{Illustration for the proof of Lemma~\ref{lem:roots}. 
    Solid lines denote any number of exact duplications and dashed 
    lines represent a mixture of exact and noisy duplications (the 
    number of noisy duplications is determined by $P_1$ and $P_2$).}
    \label{fig:1ND}
\end{figure}

The following lemma, which relates the intersection of descendant 
cones to the intersection of the sets of roots of these cones, is of 
use in the discussion of 1ND-detecting codes.
\begin{lemma}\label{lem:roots}
For any strings $x_1,x_2\in\Sigma^*$ and sets $P_1,P_2\subseteq 
\mathbb Z_{\ge0}$, 
\[\dc^{*(P_1)}(x_1)\cap \dc^{*(P_2)}(x_2)\neq\varnothing\]
if and only if 
\[\rt(\dc^{*(P_1)}(x_1))\cap \rt(\dc^{*(P_2)}(x_2))\neq\varnothing.\]
\end{lemma}
\begin{IEEEproof}
 The `only if' direction follows from definition. For the other 
 direction, assume there exist $x_1'\in \dc^{*(P_1)}(x_1)$ and 
 $x_2'\in \dc^{*(P_2)}(x_2)$ such that $\rt(x_1')=\rt(x_2')$. But 
 then, by Theorem~\ref{th:rootint}, there exists 
 $x\in\dc^{*(0)}(x'_1)\cap \dc^{*(0)}(x'_2)$. It follows that 
 $x\in\dc^{*(P_1)}(x_1)\cap \dc^{*(P_2)}(x_2)$. This is illustrated in 
 Figure~\ref{fig:1ND}, where $y=\rt(x_1')=\rt(x_2')$.
\end{IEEEproof}

We can now characterize 1ND-detecting codes in terms of duplication 
roots and descendant cones.
\begin{lemma}\label{lem:1ND-def}
A code $C\subseteq\Sigma^n$ is a 1ND-detecting $k$-duplication code if 
and only if for any two distinct codewords $c_1,c_2\in C$, 
\begin{equation}
    \dc^{*(\le1)}(c_1)\cap \dc^{*(0)}(c_2) = \varnothing, \label{eq:1ND-def}
\end{equation}
or equivalently,
\begin{align}
    \rt(c_2)&\neq\rt(c_1),\label{eq:1ND-def-1}\\ \rt(c_2)&\notin\rt(\dc^{*(1)}(c_1)).\label{eq:1ND-def-2}
\end{align}
\end{lemma}
\begin{IEEEproof}
Consider the following decoder: If there is a codeword with the same 
(exact-)duplication root as the received word, output that codeword. 
If not, declare that a noisy duplication error has occurred. Now, 
suppose~\eqref{eq:1ND-def} holds and that $c_1$ is transmitted. If 
only exact duplications occur, the decoder outputs $c_1$ since exact 
duplications do not alter the root and there is no other codeword 
$c_2$ with the same root as $c_1$. If, in addition, a noisy 
duplication occurs, then the received word either has the same root as 
$c_1$ or it does not. 
Note again that the duplication root of the received word only changes 
as a result of the noisy duplication, regardless of when it occurs in 
the sequence of duplication events.
In the former case, the decoder correctly outputs $c_1$. In the latter 
case,~\eqref{eq:1ND-def} implies that no codeword has the same root as 
the received word, and thus the decoder correctly declares that a 
noisy duplication has occurred. 

On the other hand, if~\eqref{eq:1ND-def} does not hold, no decoding 
method can both `correct any number of exact duplications' and `detect 
the presense of one noisy duplication'. That is because there exist 
distinct $c_1$ and $c_2$ and some $x\in \dc^{*(\le1)}(c_1)\cap 
\dc^{*(0)}(c_2)$. If $x$ is received then there is no way to determine 
whether $c_1$ or $c_2$ was transmitted.

The equivalence between~\eqref{eq:1ND-def} and (\ref{eq:1ND-def-1},
\ref{eq:1ND-def-2}) follows from Lemma~\ref{lem:roots}.
\end{IEEEproof}

Based on Lemma~\ref{lem:1ND-def}, we consider codes whose distinct 
codewords satisfy~\eqref{eq:1ND-def-1} and \eqref{eq:1ND-def-2}.
Further, the decoder outputs the codeword with the same root as the 
retrieved word if it exists, and otherwise declares a noisy 
duplication.

As a result of the substitution in the noisy duplication error, the 
length of the duplication root may change. One way to simplify the 
code design is to restrict ourselves to codes whose codewords all have 
duplication roots with the same length. Then, error patterns that 
modify this length can be easily detected and we can focus on patterns 
that keep the duplication-root length the same. Specifically, for a 
given length~$n$, we consider codes whose codewords are irreducible strings of length~$n$. The effect of this restriction on the size of 
the code is discussed following Theorem~\ref{thm:1SDbounds}.

\begin{definition}
A substitution error (as a component of a noisy-duplication error) 
that changes the root but not the length of the root is called an 
\emph{ambiguous substitution}.
\end{definition}

It is easy to verify that when $k=1$ a noisy duplication is never 
ambiguous. Thus, challenges arise only when $k\geq 2$. The following 
sequence of lemmas characterize the conditions under which a 
substitution is ambiguous. 

\begin{lemma}
Let $x\in\Sigma^*$ be some string resulting from a $k$-duplication, 
$k\geq 2$. If a substitution occurs (as part of a noisy duplication) 
in the last $k$ positions of $x$ then it is not ambiguous.
\end{lemma}
\begin{IEEEproof}
Since a substitution that occurs as part of a noisy duplication 
changes the copied part, we must have $z\eqdef \bar{\phi}(x)=u0^k w$, 
with $\abs{w}\leq k-1$. After the substitution we get $x'$, with 
$z'\eqdef \bar{\phi}(x')=u0^{k-i-1} b 0^{i}w$, for some $b\in\Sigma
\setminus\{0\}$ and $i+\abs{w}\leq k-1$. It is, however, obvious that 
$\abs{\mu(z)}<\abs{\mu(z')}$, and thus $\abs{\rt(x)}<\abs{\rt(x')}$.
\end{IEEEproof}

\begin{lemma}
Let $x\in\Sigma^*$ be some string resulting from a $k$-duplication, 
$k\geq 2$. If $x'$ is obtained from $x$ as a result of a substitution 
that occurs (as part of a noisy duplication) in position $\ell\leq
\abs{x}-k$, and in $\phi(x)$ positions $\ell+1,\dots,\ell+k-1$ contain 
only zeros, then the substitution is not ambiguous.
\end{lemma}

\begin{IEEEproof}
Denote $z=\bar{\phi}(x)$. Assume $z'\eqdef z+b\cdot\epsilon_{\ell-k}$ 
(where the subscript is indeed $\ell-k$ since by considering 
$\bar{\phi}(x)$ we are omitting the prefix $\hat{\phi}(x)$ of 
length~$k$). Then we may write
\[
\begingroup
\setlength\arraycolsep{2pt}
\begin{matrix}
z&=&u &0 &0^{k-1} &b' &w\\
b\cdot\epsilon_{\ell-k}&=& 0^{\abs{u}}& b& 0^{k-1}& (-b)& 0^{\abs{w}}\\
z'&=& u& b &0^{k-1}& (b'-b)& w
\end{matrix}
\endgroup
\]
where $u\in\Sigma^{\ell-k-1}$, $w\in\Sigma^*$, $b\in\Sigma\setminus
\{0\}$, and $b'\in\Sigma$. We now have two cases. If $b'\neq b$, then 
obviously $\abs{\mu(z)}<\abs{\mu(z')}$, namely $\abs{\rt(x)}<
\abs{\rt(x')}$. If $b'=b$, then $\rt(x)=\rt(x')$, which is again not 
ambiguous.
\end{IEEEproof}

The remaining cases are all handled in the following lemma.

\begin{lemma}
\label{lem:ambig}
Let $x\in\Sigma^*$ be some string resulting from a $k$-duplication, 
$k\geq 2$, and let $x'$ be obtained from $x$ as a result of a 
substitution that occurs as part of a noisy duplication. Denote 
$z\eqdef\bar{\phi}(x)$ and 
$z'\eqdef\bar{\phi}(x') = z+\epsilon_{\ell-k}$. Assume
\[
\begingroup
\setlength\arraycolsep{2pt}
\begin{matrix}
z&=&u &0^{pk+m+i-1} &0 &0^{k-i} & v & b' & w\\
b\cdot\epsilon_{\ell-k}&=& 0^{\abs{u}}&0^{pk+m+i-1}& b& 0^{k-i}& 0^{\abs{v}}& (-b)& 0^{\abs{w}}\\
z'&=&u &0^{pk+m+i-1} &b &0^{k-i} & v & (b'-b) & w\\
\end{matrix}
\endgroup
\]
where $u,w\in\Sigma^*$, $v\in\Sigma^{i-1}$, $v$ is not empty and 
begins with a non-zero letter, $b\in\Sigma\setminus\{0\}$, $b'\in
\Sigma$, the run of zeros $0^{pk+m+k}$ in $z$ between $u$ and $v$ is 
maximal, $p\in\Z_{\geq 0}$, $0\leq m < k$, $1<i\leq k$. Furthermore, 
denote the length of the run of zeros to the left of $b'$ in $z$ by 
$m_1$, and to its right by $m_2$. Then the substitution is ambiguous 
exactly when either:
\begin{enumerate}[label={C.\arabic*}]
   \item $1<i\leq k-m$, $b'=b$, and $\lfloor \frac{m_{2}}{k}\rfloor< \lfloor \frac{m_{1}+m_{2}+1}{k}\rfloor$. 
  \item $k-m<i \leq k$ and ($b'\notin\{0,b\}$ or $\lfloor \frac{m_{2}}{k}\rfloor= \lfloor \frac{m_{1}+m_{2}+1}{k}\rfloor$). 
\end{enumerate}
\end{lemma}

\begin{IEEEproof}
The following cases are possible:
\begin{enumerate}
  \item If $ 1<i\leq (k-m)$ then:
    \begin{enumerate}
      \item \label{itm:ambig2a} if $b'=b$ and $\lfloor \frac{m_{2}}{k}
      \rfloor< \lfloor \frac{m_{1}+m_{2}+1}{k}\rfloor$, then a run of 
      $0$s of length at least $k$ will be created in $z'$, leading to 
      $\abs{\mu(z')}=\abs{\mu(z)}$ but $\mu(z')\neq\mu(z)$. Thus the 
      substitution is ambiguous.
      \item if $b'=0$ and $\lfloor \frac{m_{2}}{k}\rfloor< \lfloor 
      \frac{m_{1}+m_{2}+1}{k}\rfloor$, then length of the root over 
      all increases by $2k$.
      \item in all other cases, the root's length increases by $k$.
    \end{enumerate}
  \item If $ (k-m)<i \leq k$, then a run of $0$s of length 
  $m+i-1\ge k$ will exist before $b$, implying that the length of the 
  root before $v$ will not change. Then:
    \begin{enumerate}
      \item if $b'=b$ and $\lfloor \frac{m_{2}}{k}\rfloor< \lfloor 
      \frac{m_{1}+m_{2}+1}{k}\rfloor$, then the length of the root 
      decreases by $k$.
      \item if $b'=0$ and $\lfloor \frac{m_{2}}{k}\rfloor< \lfloor 
      \frac{m_{1}+m_{2}+1}{k}\rfloor$, then the length of the root 
      increases by $k$.
      \item \label{itm:ambig3c}in all other cases, the length of the 
      root remains the same, resulting in an ambiguous substitution.
    \end{enumerate}
\end{enumerate}
\end{IEEEproof}

Examples for the two cases in which ambiguous substitutions occur, as 
described in Lemma~\ref{lem:ambig}, are given in 
Table~\ref{Table III}.
         \begin{table}
              \centering
              \caption{Examples of ambiguous substitution errors found 
              in Lemma~\ref{lem:ambig}. In all cases $y=
              \hat{\phi}(x)$, $z=\bar{\phi}(x)$, $z'=\bar{\phi}(x')$}
              \label{Table III}
              \begin{tabular}{P{4 cm}|P{4 cm}}
                \hline
                 \ref{itm:ambig2a} & \ref{itm:ambig3c}  \\ \hline
               \thead{$x=121220\underline{220}02200$\\ 
               $(y,z)=(121,102\underline{000}10201)$ \\ 
               $\rt(x)=12122002200$} & 
               \thead{$x=12122\underline{122}002200$ \\ 
               $(y,z)=(121,10\underline{000}210201)$ \\
               $\rt(x)=12122002200$ } \\ \hline
               \thead {$x'=12122022\underline{2}02200$\\ 
               $(y,z')=(121,10200\underline{2}10\underline{0}01)$ \\ 
               $\rt(x')=121220\underline{2}2200$} 
               & \thead{$x'=1212212\underline{0}002200$ \\ 
               $(y,z')=(121,1000\underline{1}21\underline{2}201)$ \\ 
               $\rt(x')=1212\underline{0}002200$} \\ \hline
              \end{tabular}
              \newline\newline
        \end{table}

\subsection{Bounds on the size of the code}
We use the analysis of the previous section to find lower bounds on 
the size of 1ND-detecting codes. For $x\in\Sigma^n$, a quantity that 
will be useful in bounding the size of codes is the following: 
\[V(x)\eqdef\abs*{\rt(\dc^{*(\le1)}(x))\cap \Sigma^n}.\] 
This counts the number of strings $x'$ that can be obtained from $x$ 
through any number of duplications, at most one of them noisy, and 
such that $\abs{\rt(x)}=\abs{\rt(x')}$. 

 \begin{lemma}\label{lem:In_ball_irr}
 For $x\in\irr(n)$, where $n \geq 2k\geq 4$,
  \begin{equation*}
     V(x)\le (n-k)(q-1)-\wt\left({\bar\phi(x)}\right)(q-2).
 \end{equation*}
\end{lemma}

\begin{IEEEproof}
We first assume, without loss of generality, that the noisy 
duplication occurs last, since subsequent duplications (which are not 
noisy) do not change the duplication root. Assume the notation is as 
defined in Lemma~\ref{lem:ambig}.

We first bound the contribution of the case \ref{itm:ambig2a} of the 
proof of Lemma~\ref{lem:ambig} to $V(x)$. Since $n\geq 2k$ and $x$ is 
irreducible, we have that $\wt(z)\geq 1$. There are $\wt(z)$ non-zero 
elements in $z$ that can serve as the first letter of $v$, which we 
shall call the \emph{anchor}. In this case, $b'\neq 0$, and it is 
found at most $k-m-1$ positions after the anchor. We contend that 
there is at most one such choice for $b'$. Indeed, if we are in 
case~\ref{itm:ambig2a}, then there is a run of $m_1$ zeros immediately 
to the left of $b'$, and $m_2$ to the right. But
\[ \floorenv*{\frac{m_1+m_2+1}{k}}>\floorenv*{\frac{m_2}{k}}\geq 0,\]
implying
\[m_1+m_2+1\geq k.\]
Thus, if case~\ref{itm:ambig2a} holds then there is a single non-zero 
element in the $k$ positions following the anchor. Additionally, since 
$b'=b$, we have a single choice for the value of $b$. Finally, we note 
that case~\ref{itm:ambig2a} cannot occur when the anchor is the last 
non-zero element in $z$. Hence, in total, the contribution of 
case~\ref{itm:ambig2a} does not exceed $\wt(z)-1$.

We now turn to the case of \ref{itm:ambig3c}. Assuming an anchor was 
chosen, the value of $i$ can take at most $m$ values, which is the 
length of the run of zeros before the anchor, taken modulo $k$. 
Ranging over all the run's zeros, the effect of modulo $k$ simply 
leaves us with a choice of a position containing a $0$ in $z$, since 
$x$ is irreducible. There are $n-k-\wt(z)$ such positions. Finally, 
there are at most $q-1$ possibilities for $b$. Thus, this case 
contributes at most $(n-k-\wt(z))(q-1)$ to $V(x)$. Noting that $x$ 
itself also contributes to $V(x)$ completes the proof.
\end{IEEEproof}

To find a lower bound on the size of the code, we apply the 
Gilbert-Varshamov (GV) bound with the average size of the sphere  
(see, e.g.,~\cite{Tol97}). 

\begin{lemma}\label{Lem:upbound_Avedeg}
Let $x$ be a randomly and uniformly chosen string from $\irr(n)$. If 
$n\geq 2k\geq 4$, then 
\[
    \mathbb E [V(x)]\le {2(n-k)(q-1)}/{q}.
\]
\end{lemma}

\begin{IEEEproof}
Let $z=\bar\phi(x)$. From Lemma~\ref{lem:In_ball_irr}, to find the 
expected value of $V(x)$, it suffices to find the expected value of 
$\wt(z)$. 

Fix $i$ and let $U$ be the set of strings obtained by removing 
position $i$ from the strings in $\rll(n-k)$ (if multiple copies of a 
string exist we keep only one). Let $S$ be the set of strings $s$ in 
$U$ that contain a run of 0s of length at least $k-1$ that includes 
$s_{i-1}$ or $s_i$. Furthermore, let $S^c=U\setminus S$. Now, the 
number of strings in $\rll(n-k)$ that contain a $0$ in position $i$ 
equals $\abs{S^c}$, while the total number of strings in $\rll(n-k)$ 
equals $\abs{S^c}q+\abs{S}(q-1)$. Hence, for a randomly chosen 
$z\in\rll(n-k)$,
\begin{equation*} \label{eq:prob_mui_0}
    \begin{split}
        \Pr(z_{i}= 0)&=\frac{\abs{S^c}}{\abs{S^c}q+\abs{S}(q-1)}\le \frac1q
    \end{split}
\end{equation*}
Thus, $\mathbb E[\wt(z)]\ge (n-k)(q-1)/q$. The result then follows 
from Lemma~\ref{lem:In_ball_irr}.
\end{IEEEproof}

The above lemma leads to the lower bound in the following theorem.

\begin{theorem}\label{thm:1SDbounds}
For positive integers $n\geq 2k\geq 4$, the maximum size 
$A_{\ond}(n,q,k)$ of a 1ND-detecting codes of length $n$ over 
$\mathbb Z_q$ satisfies
\begin{equation*}
    \frac{1}{4(n-k)}\cdot M\le A_{\ond}(n,q,k)\le M,
\end{equation*}
where 
\begin{equation}
    \label{eq:defM}
M\eqdef\sum\limits_{i=0}^{\lfloor n/k\rfloor-1} \abs*{\irr(n-ik)}
=\sum\limits_{i=1}^{\lfloor n/k\rfloor} q^k \abs*{\rll(n-ik)}
\end{equation}
is the number of irreducible words whose descendant cones  intersect 
$\Sigma^n$.
\end{theorem}

\begin{IEEEproof}
First we show that
\begin{equation*}
     \frac{q^{k+1} \abs*{\rll(n-k)}}{2(n-k)(q-1)}\! 
     \le A_{\ond}(n,q,k)\le M.
\end{equation*}
The lower bound follows by applying the generalized GV 
bound~\cite{Tol97} with Lemma~\ref{Lem:upbound_Avedeg}. The upper 
bound follows from the fact that the code must be able to correct any 
number of duplication errors and from~\cite{jain2017e} where such 
codes are discussed.

To get the lower bound to the more appealing form we claim, we note 
that to any string of length $m-k$ that has no $0^k$ substring, we can 
append a string of length $k$ whose first element is nonzero, and thus 
obtain a string of length $m$ that has no $0^k$ substring. Hence, 
\[\abs*{\rll(m)} \ge \abs*{\rll(m-k)} (q-1)q^{k-1}.\]
Thus
\[
\abs*{\rll(n-ik)}
\le\frac{\abs*{\rll(n-k)}}{(q-1)^{i-1}q^{(i-1)(k-1)}}.
\]
We then have
\begin{align*}
M & =\sum_{i=1}^{\lfloor n/k\rfloor}q^{k}\abs*{\rll(n-ik)}\\
 & \le q^{k}\abs*{\rll(n-k)}\sum_{i=1}^{\lfloor n/k\rfloor}\frac{1}{(q-1)^{i-1}q^{(i-1)(k-1)}}\\
 & \le q^{k}\abs*{\rll(n-k)}\sum_{i=1}^{\infty}\frac{1}{(q-1)^{i-1}q^{(i-1)(k-1)}}\\
 & \le q^{k}\abs*{\rll(n-k)}\frac{(q-1)q^{k-1}}{(q-1)q^{k-1}-1}.
\end{align*}
Since $q+k\ge 4$, 
\begin{equation}\label{eq:contRootLength}
    \abs{\irr(n)}=q^k \abs*{\rll(n-k)}\ge M/2,
\end{equation}
and we have the desired claim.
\end{IEEEproof}

\subsection{Code construction}

The goal of this section is to construct 1ND-detecting codes. We shall 
first consider an auxiliary code construction which will be useful not 
only here, but also in the following section. The error we would like 
to detect by this auxiliary code is as follows:

\begin{definition}
For $n,k>0$, let $z,z'\in\Sigma^n$ be some strings. If we can write
\[
\begingroup
\setlength\arraycolsep{2pt}
\begin{matrix}
    z&=&u&v&w&0^{\abs{v}}&x\\
    z'&=&u&0^{\abs{v}}&w&v&x
\end{matrix}
\endgroup
\]
where $u,v,w,x\in\Sigma^*$, $1\leq \abs{v}\leq k-1$, $v$ is a non-zero 
string, and $\abs{v}+\abs{w}=k$, then we say $z$ and $z'$ differ by a 
single $k$-switch error.
\end{definition}

Intuitively, a single $k$-switch error takes a non-zero non-empty 
substring of length at most $k-1$, and switches it with an all-zero 
substring of the same length found $k$ positions before or after it.

Any non-empty string $z\in\Sigma^n$ may be partitioned into 
non-overlapping blocks of length $k$:
\[ z= B_1(z) B_2(z) \dots B_{\ceilenv{n/k}}(z),\]
where $B_i(z)\in\Sigma^k$ for all $i$, except if $k$ does not divide 
$n$, in which case, $B_{\ceilenv{n/k}}\in\Sigma^{n\bmod k}$. We note 
that $k$ is implicit in the definition of $B_i(z)$.

We now give a construction for a family of codes which we then show 
are all capable of detecting a single $k$-switch error.

\begin{construction}\label{Const:auxcode}
Let $k\geq 2$ and let $p$ be the smallest odd integer larger than 
$k-1$, namely
\[ p\eqdef 2\ceilenv*{\frac{k-1}{2}}+1.\]
Fix a code length $n\in\N$ and let $S\subseteq\Sigma^n$ be an 
arbitrary set of strings. For any string $x\in S$, and $\ell=0,1,2,3$, 
we define 
\[ Z_\ell(x)\eqdef \sum_{i\in I_\ell} \abs{B_{i}(x)}_0,\]
where $I_\ell=\mathset*{1\leq t\leq \ceilenv{n/k}}{t\equiv\ell 
\pmod 4}$. For all \linebreak$0\le i,j<p$, we construct
\begin{align*}
  C^{\aux}_{i,j}(S) \eqdef \big\{x\in S ~\big|~ Z_0(x)+2Z_2(x) &\equiv i \pmod{p},\\Z_1(x)+2Z_3(x) &\equiv j \pmod{p}\big\}.
\end{align*}
\end{construction}

\begin{theorem}\label{th:auxcode}
Each code $C^{\aux}_{i,j}(S)$ of Construction~\ref{Const:auxcode} can 
detect a single $k$-switch error or a single zero replaced by a 
non-zero letter.
\end{theorem}

\begin{IEEEproof} 
Since $k\geq 2$ we have $p\geq 3$ which immediately enables the 
detection of a single zero replaced by a non-zero letter. Let us 
therefore focus on the problem of detecting a single $k$-switch error.

We assume $n\geq k+1$, otherwise the claim is trivial. Assume 
$x\in C^{\aux}_{i,j}(S)$ sustains a single $k$-switch error, resulting 
in the string $x'\in\Sigma^n$. For $0\le \ell\le 3$, let 
\[\Delta_\ell\eqdef Z_\ell(x')-Z_\ell(x).\]
Furthermore, for $0\le \ell \le 1$, let 
\[F_\ell\eqdef \Delta_{\ell}+2\Delta_{\ell+2}.\]
To prove the error detection capabilities of the code it now suffices 
to show that
\begin{equation}\label{eq:aux_det_indi_code}
    F_0\not\equiv 0 \pmod{p}\qquad\text{ or }\qquad F_1 \not\equiv 0 \pmod{p}.
\end{equation}

Based on the definition of a $k$-switch error, the number of zeros 
changes in some blocks. We consider the following possible cases.

First, if the number of zeros changes in 2 consecutive blocks, then 
one of the pairs $(\Delta_0, \Delta_1)$, $(\Delta_1, \Delta_2)$, 
$(\Delta_2, \Delta_3)$, $(\Delta_3, \Delta_0)$ equals 
$(\delta,-\delta)$ for $0<|\delta|<k$, and the two other $\Delta$'s 
are equal to $0$. Then, $|F_0|=|\delta|$ or $|F_0|=2|\delta|.$ In the 
former case $F_0\not\equiv 0\pmod{p}$ since $0<|\delta|< k\le p$. In 
the latter case, $F_0\not\equiv 0\pmod{p}$ since $0<2|\delta|<2p$ and 
$2\delta\neq p$ (recall that $p$ is odd).

Second, if the number of zeros changes in two non-consecutive blocks, 
then only one of the pairs $(\Delta_0, \Delta_2)$ and 
$(\Delta_1, \Delta_3)$ equals $(\delta,-\delta)$ for $0<|\delta|<k$, 
and the other equals $(0,0)$. Then, either  $|F_0|=|\delta|$ or 
$|F_1|=|\delta|$, and in both cases~\eqref{eq:aux_det_indi_code} is 
satisfied.

Third, if the change of number of zeros occurs in three consecutive 
blocks, then there exists $\ell$ such that $\Delta_\ell=\delta'\neq 0$ 
and $\Delta_{2+\ell}=0$ (indices taken modulo $4$), where $0<|\delta'|
<k$ and $2|\delta'|\neq p$. Then either $F_0$ or $F_1$ takes on the 
value of $\delta'$ or $2\delta'$. But $\delta'\not\equiv 0\pmod{p}$ 
and $2\delta'\not\equiv 0\pmod{p}$, implying that 
\eqref{eq:aux_det_indi_code} is satisfied.
\end{IEEEproof}

We now turn to construct 1ND-detecting codes. As before, we consider 
codes that consist of irreducible strings of length $n$. We thus need 
to devise a method to detect ambiguous substitutions.

As mentioned before, when $k=1$ ambiguous substitutions cannot occur. 
Hence $\irr(n)$ is a 1ND-detecting code. For \linebreak$k\geq 2$, our 
analysis rests on the following lemma.

\begin{lemma}
\label{lem:ambiswitch}
Let $k\geq 2$. If $x\in\Sigma^*$ and $x'$ is obtained from $x$ via any 
number of duplications among which one contains an ambiguous 
substitution, then $\bar\phi(\rt(x))$ and $\bar\phi(\rt(x'))$ differ 
by a single $k$-switch error, or
\[ \abs*{\abs{\bar\phi(\rt(x))}_0-\abs{\bar\phi(\rt(x'))}_0}=1. \]
\end{lemma}
\begin{IEEEproof}
Denote $z\eqdef\bar\phi(x)$ and $z'\eqdef\bar\phi(x')$. With the 
notation of Lemma~\ref{lem:ambig}, one can verify that in 
Case~\ref{itm:ambig2a} we have
\begin{equation}
\label{eq:ambig2a}
\begingroup
\setlength\arraycolsep{2pt}
\begin{matrix}
    \mu(z)&=&u'& v &0^{i-1-\abs{v}}b 0^{k-i}&0^{\abs{v}}&w'
    \\
    \mu(z')&=&u'& 0^{\abs{v}} &0^{i-1-\abs{v}}b0^{k-i}&v &w'
\end{matrix}
\endgroup
\end{equation}
and in Case \ref{itm:ambig3c},
\begin{equation}
\label{eq:ambig3c}
\begingroup
\setlength\arraycolsep{2pt}
\begin{matrix}
        \mu(z)&=&u' &0 &0^{k-\abs{v}-1} v &b'  &w'\\
        \mu(z')&=&u'&b &0^{k-\abs{v}-1}v &(b'-b) &w'
\end{matrix}
\endgroup
\end{equation}
for some $u',w'\in \Sigma^*$. In~\eqref{eq:ambig2a} we see a single 
$k$-switch error. In~\eqref{eq:ambig3c}, if $b'=b$ we have a single 
$k$-switch error, and if $b\neq b'$ then the number of zeros differ by 
one.
\end{IEEEproof}

\begin{construction}\label{Const:code}
Let $n,k$ be positive integers, $n\geq k$, and let $S\eqdef\rll(n-k)$. 
For all $0\leq i,j<p$, we construct
\[ C_{i,j}\eqdef \mathset*{\phi^{-1}(yz)}{y\in\Sigma^k, z\in C^{\aux}_{i,j}(S)},\]
where $p$ and $C^{\aux}_{i,j}(S)$ are defined in 
Construction~\ref{Const:auxcode}.
\end{construction}

\begin{theorem}\label{th:error_det_code}
With the setting as in Construction~\ref{Const:code}, the code 
$C_{i,j}$ is a 1ND-detecting code. \end{theorem} 

\begin{IEEEproof}
By our choice of $S$, we necessarily have that $C_{i,j}\subseteq
\irr(n)$. If $k=1$, then $C_{0,0}=\irr(n)$ is the only code and the 
theorem is immediate. 

Assume $k\geq 2$. Let $c_1,c_2\in C_{i,j}$ be distinct codewords. 
Since $C_{i,j}\subseteq\irr(n)$, $\rt(c_1)=c_1$ and $\rt(c_2)=c_2$, 
which are distinct. Based on~\eqref{eq:1ND-def-2} it suffices to show 
that for any $c'_1\in \dc^{*(1)}(c_1)$, we have $c_2\neq\rt(c'_1)$.

If $\rt(c'_1)=\rt(c_1)=c_1$, then clearly $c_2\neq \rt(c'_1)$. So we 
assume $\rt(c'_1)\neq c_1$. It is then sufficient to show that 
$\rt(c'_1)\notin C_{i,j}$. This is obvious if $|\rt(c'_1)|\neq n$ and 
the substitution is not ambiguous. If the substitution is ambiguous, 
we obtain the claimed result by combining Lemma~\ref{lem:ambiswitch} 
and Theorem~\ref{th:auxcode}.
\end{IEEEproof}

\begin{corollary}
\label{cor:psquared}
If $n\geq k\geq 2$ then
\[ A_{\ond}(n,q,k) \geq \frac{1}{2(k+1)^{2}}\cdot M,\]
where $M$ is given by~\eqref{eq:defM}.
\end{corollary}
\begin{IEEEproof}
Let $p$ and $C_{i,j}$ be defined as in Construction~\ref{Const:code}. 
The set $\mathset{ C_{i,j}}{0\leq i,j <p}$ forms a partition of 
$\irr(n)$. Thus, a simple averaging argument shows that there exist 
$i$ and $j$ such that
\[ \abs{C_{i,j}} \geq \frac{\abs{\irr(n)}}{p^2}.\]
Since $p\leq k+1$, and by~\eqref{eq:contRootLength}, we obtain the 
claim.
\end{IEEEproof}

Note that the lower bound on $A_{\ond}(n,q,k)$ in this corollary may 
be better than the one given in Theorem~\ref{thm:1SDbounds}.

The problem with the bound of Corollary~\ref{cor:psquared} is that it 
is not constructive. In particular, we do not know exactly what choice 
of $i$ and $j$ gives the largest code $C_{i,j}$ in 
Construction~\ref{Const:code}. \autoref{Const:codeconstructive} below 
provides a sub-code of $C_{0,0}$ from \autoref{Const:code} whose size 
can be lower bounded, albeit, somewhat smaller than the guarantee of 
Corollary~\ref{cor:psquared}.

\begin{construction}\label{Const:codeconstructive}
Let $k\geq 2$ and let $p$ be the smallest odd integer larger than 
$k-1$, namely
\[ p\eqdef 2\ceilenv*{\frac{k-1}{2}}+1.\]
Fix a code length $n\in\N$, $n\geq 5k$. We construct a code 
$C\subseteq\Sigma^n$ in the following way: For each $y\in\rll(n-5k)$, 
construct four strings of length $k$, denoted $B_0,B_1,B_2,B_3
\in\Sigma^k$,
\[
    B_i = 0^{\beta_i} 1^{k-\beta_i}, \qquad\forall 0\leq i\leq 3
\]
where
\begin{align*}
    \beta_i & = (-(\zeta_i+2\zeta_{i+2})\bmod p)-2\beta_{i+2}, & i&=0,1\\
    \beta_{i+2} & = \floorenv*{\frac{(-(\zeta_i+2\zeta_{i+2})\bmod p)}{2}}, & i&=0,1\\
    \zeta_i &= Z_i(\phi^{-1}(0^k y)), &
    i&=0,1,2,3
\end{align*}
and add the codewords $\phi^{-1}(B B_0 B_1 B_2 B_3 y)$ where $B$ runs 
over all strings in $\Sigma^k$.
\end{construction}

\begin{theorem}
Let $q$ be the alphabet size, $k$ the duplication length, $q+k\geq 4$, 
and $n\in\N$, $n\geq 5k$. Then the code $C$ from 
Construction~\ref{Const:codeconstructive} is a 1ND-detecting code of 
size
\[ \abs{C} = \irr(n-4k)\geq \frac{1}{2\cdot q^{4k}}\cdot M,\]
where $M$ is given in~\eqref{eq:defM}.
\end{theorem}
\begin{IEEEproof}
One can easily verify that $0\leq \beta_1 < k$, hence all the blocks 
$B_i$ end with a non-zero symbol and therefore all the codewords are 
irreducible. Additionally, by inspection we can verify that 
$C\subseteq C_{0,0}$, where $C_{0,0}$ is obtained from 
Construction~\ref{Const:code}. Thus, $C$ is 1ND-detecting. Finally, 
all the codewords constructed are distinct, hence 
\[\abs{C}=q^k\abs*{\rll(n-5k)}=\abs*{\irr(n-4k)}\geq 
\frac{1}{2\cdot q^{4k}}\cdot M,\]
where the last inequality follows from the fact that 
$\abs*{\irr(n-4k)}\geq \abs*{\irr(n)}/q^{4k}$ and then 
from~\eqref{eq:contRootLength}.
\end{IEEEproof}

%%%%%%%%%%%%%%%%%%%%%%%%%%%%%%%%%%%%%%%%%%%%%%%%%%%%%%%%%%%%%%%%%%%%%%
\section{Unrestricted Error-detecting codes}
\label{sec:detect_un_subst}
%%%%%%%%%%%%%%%%%%%%%%%%%%%%%%%%%%%%%%%%%%%%%%%%%%%%%%%%%%%%%%%%%%%%%%
Substitution mutations might occur not only in duplication copies, but 
also independently in other positions. In what follows, we consider a 
single substitution error occurring in addition to however many 
duplications, at any stage during the sequence of duplication events,
but not necessarily in a duplicated substring. We refer to codes 
correcting many duplication errors and detecting a single independent 
substitution error as \emph{1S-detecting} codes.
 
 \begin{definition}
A code $C\subseteq\Sigma^*$ is a 1S-detecting code if there exists a 
decoding function $\cD:\Sigma^*\to C\cup\{\mathrm{error}\}$ such that 
if $c\in C$ was transmitted and $y\in\Sigma^*$ was received then 
$\cD(y)=c$ if only duplication errors occurred, and $\cD(y)\in
\{c,\mathrm{error}\}$ if in addition to the duplications, exactly one 
unrestricted substitution occurred.
\end{definition}

 \begin{lemma}
 \label{lem:1scond}
 A code $C\in \Sigma^{n}$ is a 1S-detecting code if and only if for 
 any two distinct codewords $c_1,c_2\in C$, we have 
 \begin{equation}
 \label{eq:1scond}
 \rt(c_1)\neq \rt(c_2) \quad\text{and}\quad \rt(c_2)\notin\rt(\dc^{*,1}(c_1)).
\end{equation}
\end{lemma}
 
 \begin{IEEEproof}
 In the one direction, we define for any $y\in\Sigma^*$, $\cD(y)=c$ if 
 $\rt(c)=\rt(y)$, and $\cD(y)=\mathrm{error}$ otherwise. Clearly 
 if~\eqref{eq:1scond} holds then $\cD$ is a decoding function proving 
 that $C$ is a 1S-detecting code.
 
 In the other direction, if~\eqref{eq:1ND-def} does not hold we have 
 two (not mutually exclusive) cases. If there exist $c_1,c_2\in C$ 
 such that $\rt(c_1)=\rt(c_2)$ then by Theorem~\ref{th:rootint} there 
 exists $y\in \dc^{*,0}(c_1)\cap \dc^{*,0}(c_2)$ and no decoding 
 function can always correctly decode $y$. Similarly, if $\rt(c_2)\in 
 \rt(\dc^{*,1}(c_1))$ then there exists $y\in \dc^{*,1}(c_1)$ such 
 that $\rt(y)=\rt(c_2)$ and no decoding function $\cD$ can always 
 decode $y$ correctly.
 \end{IEEEproof}

We shall adopt the same general strategy as the previous section. 
Namely, we will construct a code based on irreducible words of 
length ~$n$. Descendants whose duplication root is not of length~$n$ 
will be easily detected. Our challenge is therefore to detect errors 
that do not change the length of the root caused by, what we refer to 
as, ambiguous substitutions.

\begin{definition}
An unrestricted substitution error that changes the root but not the 
length of the root is called an \emph{ambiguous unrestricted 
substitution}.
\end{definition}

As in the previous section, when the duplication length is $k=1$ there 
are no ambiguous unrestricted substitutions. In that case $\irr(n)$ 
can easily serve as a 1S-detecting code. Thus, we shall focus on the 
case of $k\geq 2$.

\begin{lemma}\label{lem:multi_ambi_subti_cases}
Let $n\geq 2k\geq 4$. For any string $x\in \Sigma^{n}$, let $x'\in 
\rt(\dc^{*,\leq1}(x))\cap \Sigma^{n}$ be a string obtained from $x$ 
via a single ambiguous unrestricted substitution. If
\[ d(\phi(\rt(x)),\phi(\rt(x')))\geq 3,\]
then $\bar\phi(\rt(x))$ and $\bar\phi(\rt(x'))$ differ by a single 
$k$-switch error.
\end{lemma}

To improve the flow of reading, the proof of this technical lemma is 
given in the appendix.

Our strategy, based on Lemma~\ref{lem:multi_ambi_subti_cases}, is to 
build a code as an intersection of two other component codes. If one 
component code can detect the swapping of two substrings and the other 
component code has a minimum Hamming distance of $3$ or more, then 
their intersection is a 1S-detecting code.

\begin{construction}
\label{Const:1sdetcode}
Let $q$ be a prime power, and $\Sigma\eqdef\F_q$ be the finite field 
of $q$ elements. Let $n\geq k\geq 2$ and let $r$ be the unique 
positive integer such that $\frac{q^{r-1}-1}{q-1}< n\leq 
\frac{q^r-1}{q-1}$, namely,
\[ r\eqdef \ceilenv*{\log_q (n(q-1)+1)}.\]
Denote by $C^H$ the $[n,n-r,3]$ shortened Hamming code over $\F_q$, 
and by $C^H_0,C^H_1,\dots,C^H_{q^r-1}$ its $q^r$ cosets. Finally, let 
$p$ and $C_{i,j}$ be defined as in Construction~\ref{Const:code}. For 
all $0\leq i,j<p$ and $0\leq \ell < q^r$, we construct
\[
C_{i,j,\ell}\eqdef \mathset*{c\in C_{i,j}}{\phi(c)\in C^H_\ell}.
\]
\end{construction}

\begin{theorem}
With the setting as in Construction~\ref{Const:1sdetcode}, the code 
$C_{i,j,\ell}$ is a 1S-detecting code. In particular, there exist 
$i,j,\ell$ such that
\[\abs{C_{i,j,\ell}}\geq \frac{\abs*{\irr(n)}}{q^r p^2}
\geq \frac{\abs*{\irr(n)}}{q(n(q-1)+1)(k+1)^2}.\]
\end{theorem}

\begin{IEEEproof}
By Construction~\ref{Const:code} we have that $C_{i,j}\subseteq\irr(n)$, 
hence also $C_{i,j,\ell}\subseteq\irr(n)$, which implies it can 
correct any number of duplications. Thus, following 
Lemma~\ref{lem:1scond}, it only remains to consider two distinct 
codewords $c_1,c_2\in C_{i,j,\ell}$ and show that $\rt(c_2)
\notin\rt(\dc^{*,1}(c_1))$, namely consider the case in which a single 
ambiguous unrestricted substitution occurred as part of the 
duplications.

Assume to the contrary this is not the case. By 
Lemma~\ref{lem:multi_ambi_subti_cases}, if $d(\phi(c_1),\phi(c_2))
\geq 3$, then $\bar\phi(c_1)$ and $\bar\phi(c_2)$ differ by a single 
$k$-switch error, and this is contradicts the fact that $C_{i,j}$ 
detects a single $k$-switch error in the $\bar\phi$ part of the root, 
a fact that has already been used in Theorem~\ref{th:error_det_code}. 
If $d(\phi(c_1),\phi(c_2))\leq 2$, then this contradicts the minimum 
distance implied by using the shortened Hamming code.

Finally, the existence of the code with the lower bounded size is 
guaranteed using a simple averaging argument since 
$\mathset{C_{i,j,\ell}}{0\leq i,j<p,0\leq \ell < q^r}$ forms a 
partition of $\irr(n)$.
\end{IEEEproof}

%%%%%%%%%%%%%%%%%%%%%%%%%%%%%%%%%%%%%%%%%%%%%%%%%%%%%%%%%%%%%%%%%%%%%%
\section{Unrestricted Error-Correcting Codes}
\label{sec:correct}
%%%%%%%%%%%%%%%%%%%%%%%%%%%%%%%%%%%%%%%%%%%%%%%%%%%%%%%%%%%%%%%%%%%%%%

In this section, we again observe the case of many tandem-duplications 
and a single substitution, occurring at any point during the 
duplication sequence, and not necessarily in a duplicated substring. 
However, unlike previous sections, there is no mix of correction and 
detection -- rather we aim to \emph{correct} all duplications and a 
single substitution (occurring at any stage during the sequence of 
duplication events), which makes the definition of the code more 
straightforward. We refer to codes able to correct such errors as a 
\emph{single-substitution correcting} (1S-correcting) code. Obviously, 
a code $C$ is 1S-correcting if and only if for any two distinct 
codewords $c_1,c_2\in C$, we have
\begin{equation*}
    \dc^{*,\leq 1}(c_1)\cap \dc^{*,\leq 1}(c_2) = \varnothing.
\end{equation*}

In this context, we will find it easier to consider strings in the 
$\phi$-transform domain. 
We also define the \emph{substitution distance} $\sigma(u,v)$ to 
measure the number of substitutions required to transform one string 
into the other, when $u,v$ are assumed to be in the transform domain. 
More precisely, if $u,v\in \Sigma^n$ and $v-u = \sum_{i=1}^n 
a_i \cdot\epsilon_i$, then 
\[\sigma(u,v) \deq \abs*{\mathset*{1\leq i\leq n}{a_i\neq 0}}.\]

\subsection{Error-correcting codes}

In contrast to \autoref{lem:multi_ambi_subti_cases} and 
\autoref{Const:1sdetcode}, we shall see in the following example that 
an intersection of a single substitution correcting code with a 
duplication correcting code is not, in general, a 1S-correcting code.
\begin{example}\label{ex:tan-mut-ambiguity}
Set $\Sigma=\Z_2$ and $k=3$, and observe the following two sequences 
of duplication and substitution, as seen in the $\phi$-transform 
domain:
\begin{align*}
u \deq 111010111 \to 111010111\underline{000} 
\to 1110\underline{0}01\underline{0}1000 \\
v \deq 111101010 \to 111\underline{000}101010 
\to 1110001010\underline{0}0
\end{align*}

It is clear that if $C\subseteq\Sigma^{\geq k}$ is a code correcting 
even a single duplication and a single substitution, even given the 
order in which they occur, then $\phi^{-1}(u)=111101010$ and 
$\phi^{-1}(v)=111010000$ cannot both belong to $C$. 
Observing that $u,v\in\rll(9)$ and $\sigma(u,v)=4$, however, we find 
that $C\deq\bracenv{\phi^{-1}(u),\phi^{-1}(v)}$ can correct any number 
of duplications, or correct a single substitution. 
Simple intersections, hence, do not suffice for a code correcting 
a combination of such errors.
\end{example}

In what follows, we propose a constrained-coding approach which 
resolves the issue demonstrated in the last example. It relies on the 
following observation: substitution noise might create a $0^k$ 
substring in the transform domain--that is not due to a 
duplication--as well as break a run of zeros.
However, a constrained system exists which allows us to de-couple the 
effects of duplication and substitution noise.

More precisely, we denote 
\[
\cW \eqdef \mathset*{u\in\Sigma^{\geq k}}{\forall\ \text{substring}\ v\ \text{of}\ u, |v|=k: \wt(v)>1}.
\]
We shall show that intersecting a single-substitution-error-correcting 
code with the reverse image of $\cW\cap\Sigma^{n-k}$, instead of 
$\rll(n-k)$, is a 1S-correcting code. More precisely, we aim to show 
that restricting codewords to be taken from $\cW$ (in the transform 
domain), the following holds.

\begin{lemma}\label{lem:zero-dedup}

Take an irreducible $x\in\Sigma^{\geq k}$, and $y\in\dc^{*,\leq 1}(x)$. 
If $v\deq \bar{\phi}(y)$ contains a $0^k$ substring, and $\bar{v}$ is 
derived from $v$ by removing that substring, and if $\bar{\phi}(x)
\in\cW$, then $\bar{v}\in \bar{\phi}\parenv*{\dc^{*,\leq 1}(x)}$.
\end{lemma}
\begin{IEEEproof}
We denote 
\[v = \alpha c 0^k \beta\]
for $0\neq c\in\Sigma$ and $\alpha,\beta\in \Sigma^*$, and by abuse of 
notation assume $\abs*{\alpha c} \geq 0$ is the shortest with the 
properties stated above (allowing $v = 0^k \beta$ as a private case).

We also take $y'\in \dc^{*,0}(x)$ to be the descendant of $x$ derived 
by the same sequence of duplications as $y$, where a substitution 
never occurs, and 
\[
v' = \bar{\phi}(y') = \alpha' c' 0^j a 0^{k-j-1} \beta' ,
\]
for $0\leq j<k$, $c',a\in\Sigma$, $\alpha',\beta'\in\Sigma^*$, where 
$\abs*{\alpha' c'} = \abs*{\alpha c}$. 
(We know $v'$ can be represented in this fashion since $y$ suffered a 
single substitution.)

If $a = 0$ then the claim is trivial. Assume, therefore, $a\neq 0$. 
Note that $\bar{\phi}(x) \in \cW$ and $\wt_H(0^j a 0^{k-j-1}) = 1$, 
implying that $0^{k-j-1} \beta'$ begins with a $k$-tuple of zeros. 
I.e., $\beta' = 0^{j+1} \beta''$, for some $\beta''\in\Sigma^*$. Thus, 
a descendant of $x$ is also $z'$, where $\bar{\phi}(z') 
= \alpha' c' 0^j a \beta''$.

We now reexamine $v',v$:
\[
\begingroup
\setlength\arraycolsep{2pt}
\begin{matrix}
v' &=& \alpha' & c' & 0^j & a & 0^{k-j-1} &\beta' \\
v &=&  \alpha  & c  & 0^j & 0 & 0^{k-j-1} &\beta
\end{matrix}
\endgroup
\]
and since $y$ is derived from $x$ by the same sequence of 
tandem-duplications as $y'$, with a single substitution, we may 
deduce that $\alpha,\beta$ and $\alpha',\beta'$ differ, respectively, 
in precisely one of the following manners:
\begin{itemize}
\item
There exist $b\in\Sigma$ and $\alpha_1,\alpha_2\in\Sigma^*$, with 
$\abs*{\alpha_2 c}=k-j-1$, such that
\[
\begingroup
\setlength\arraycolsep{2pt}
\begin{matrix}
v' &=& \alpha_1 &(b-a) &\alpha_2 &c' &0^j &a & 0^{k-j-1}\beta \\
v &=& \alpha_1 &b      &\alpha_2 &c  &0^j &0 & 0^{k-j-1}\beta
\end{matrix}
\endgroup
\]
and, again, by abuse of notation, including the case of 
$\abs*{\alpha_2 c}=0$, meaning $b=c$ and $b-a=c'$; in all other cases 
$c'=c$.

In this case
\begin{align*}
\bar{v} &= \alpha c \beta = \alpha_1 b \alpha_2 c 0^{j+1} \beta'' \\
&= \alpha_1 (b-a) \alpha_2 c' 0^j a \beta'' 
+ a\cdot \epsilon_{\abs*{x c}+j-k} \\
&= \bar{\phi}(z) + a\cdot \epsilon_{\abs*{x c}+j-k} .
\end{align*}

\item
$\beta = 0^j a \beta''$, implying $\alpha' c' = \alpha c$ and 
\begin{align*}
\bar{v} &= \alpha c \beta = \alpha c 0^j a \beta'' = \bar{\phi}(z) .
\end{align*}

\item
There exist $s\geq 0$, $b\in\Sigma$, $\gamma\in \Sigma^{k-1}$ and 
$\beta'''\in\Sigma^*$ such that $\beta'' = 0^{sk} \gamma b \beta'''$, 
and 
\[
\begingroup
\setlength\arraycolsep{2pt}
\begin{matrix}
v' &=& \alpha c &0^j & a & 0^{k-j-1} &0^{j+1+sk} \gamma & b & \beta''' \\
v &=&  \alpha c &0^j & 0 & 0^{k-j-1} &0^{j+1+sk} \gamma & (b+a) & \beta'''
\end{matrix}
\endgroup
\]
Let $z''$ be the ancestor of $z'$ (thus descendant of $x$) satisfying 
\[
\bar{\phi}(z'') = \alpha c 0^j a \gamma b \beta'''
\]
and note that 
\begin{align*}
\bar{v} &= \alpha c 0^{j+sk} 0 \gamma (b+a) \beta''' \\
&= \alpha c 0^{j+sk} a \gamma b \beta''' + (-a) 
\cdot \epsilon_{\abs*{\alpha c} + sk + j} \\
&\in \bar{\phi}\parenv*{\dc^{*,0}\parenv*{z'' + (-a) 
\cdot e_{\abs*{\alpha c} + j}}}
\end{align*}
\end{itemize}
\end{IEEEproof}

Recall from \cite{jain2017e} that a decoder for correcting an 
unbounded number of duplications simply has to remove incidents of 
$0^k$ from the $\bar\phi$-part of the noisy string. This lemma shows 
that the same approach can be taken with the addition of a single 
substitution--without increasing the substitution distance--provided 
that coding is done in $\cW$.

Next, we consider the case where a substitution breaks a run of zeros 
(in the transform domain).
The following lemma allows us to remove appearances of $0^ja0^{k-1-j}$ 
from the $\bar\phi$-part of a noisy string (by applying an appropriate 
substitution) without increasing the substitution distance. 
\begin{lemma}\label{lem:wrong-dedup}

Suppose $u\in\Sigma^{\geq k}$ contains a substring $0^k$ starting at 
index $i$, and suppose $v = u+a\cdot\epsilon_\ell$ for some $i\leq 
j<i+k$, $0\neq a\in\Sigma$, and $\ell\in\bracenv*{j,j-k}$ (so that 
$v_j\neq 0$). 
Note that $v' \deq v - v_j \cdot\epsilon_j$ has a $0^k$ substring at 
index $i$ (like $u$); we remove that substring from both $u,v'$ to 
produce $\bar{u},\bar{v}$, respectively. Then, irrespective of what 
value $\ell$ takes, $\sigma(\bar{u},\bar{v})\leq 1$.
\end{lemma}
\begin{IEEEproof}
The lemma is straightforward to prove by case for $\ell$. If $\ell=j$ 
then $v'=u$, and consequently $\bar{v}=\bar{u}$.

Otherwise, $\ell=j-k$ and $v_j = -a$, hence 
\[v' = u + a\cdot\parenv*{\epsilon_{j-k} + \epsilon_j}\]
and $\bar{v} = \bar{u} + a\cdot\epsilon_{j-k}$, which concludes the 
proof.
\end{IEEEproof}

It is therefore seen that a restriction to $\cW$ allows the correction 
of the substitution error without encountering the issue demonstrated 
in \autoref{ex:tan-mut-ambiguity}. This fact is more precisely stated 
in the following theorem:

\begin{theorem}\label{cor:tan-sing_mut}
If $C\subseteq\Sigma^{n}$, $n\geq k$, is an error-correcting code for 
a single substitution, and $\bar\phi(C)\subseteq\cW$, then $C$ is a 
1S-correcting code.
\end{theorem}
\begin{IEEEproof}
Take $x\in C$, $y\in \dc^{*,\leq 1}(x)$, and define $u\eqdef 
\hat\phi(x)$, $v \eqdef \bar{\phi}(y)$. We first remove $0^k$ 
substrings from $v$, stopping if we reach length~$n-k$. By 
\autoref{lem:zero-dedup}, every removal of $0^k$ does not increase the 
substitution distance of the received sequence from a duplication 
descendant of $x$; if indeed it is possible to arrive at $\hat{v}$ of 
length~$n-k$, then the error-correcting capabilities of $C$ now 
suffice to deduce $x$ from $\phi^{-1}(u\hat{v})$.

The only other possible case is that we ultimately arrive at $\hat{v}$ 
of length~$n$ which contains a substring of length~$k$ of weight~$1$. 
We remove that substring to obtain $\hat{v}'$, and reverse the 
$\phi$-transform, namely, $y'\eqdef\phi^{-1}(u\hat{v}')$. By 
\autoref{lem:wrong-dedup}, this produces $y'$ of the same length as 
$x$ and differing from it by at most a single substitution, which we 
may once more correct in the standard fashion.
\end{IEEEproof}

\subsection{Code Construction and Size}

In this section we construct a family of codes satisfying 
\autoref{cor:tan-sing_mut}. We also study the 
redundancy and rate of the proposed construction. 
We start by bounding the rate loss of using constrained coding by 
restricting codes to $\cW$:

\begin{lemma}\label{lem:redun}
For every integers $q\geq 2$ and $n\geq k\geq 1$, 
\[
\frac{r(\cW\cap\Sigma^n)}{n} \leq \frac{2}{k} \log_q\frac{q}{q-1}.
\]
\end{lemma}
\begin{IEEEproof}
We note that 
$C_n \subseteq \cW\cap\Sigma^n,$
where $C_n$ is the set of length-$n$ strings in which, 
divided into blocks of length $k$, every block ends with 
two non-zero elements. Hence, 
\begin{align*}
\frac{r(\cW\cap\Z_q^n)}{n} &\leq \frac{r(C_n)}{n} = 
\frac{1}{n}\parenv*{\floorenv*{\frac{n}{k}}+\floorenv*{\frac{n+1}{k}}} \\
&\leq \frac{2}{k} \log_q\frac{q}{q-1}.
\end{align*}
\end{IEEEproof}

\begin{theorem}
\label{thm:ecc}
If $q$ is a prime power, $r\geq 2$, and $n = {\frac{q^r-1}{q-1} 
+ \ceilenv*{\frac{2r}{k}}}$, then a 1S-correcting $k$-duplication code 
$C\subseteq\cW\cap\F_q^n$ exists, with 
\[
R(C) \geq 1 - \frac{2}{k} \log_q\frac{q}{q-1} - o(1).
\]
\end{theorem}
\begin{IEEEproof}
We begin by encoding data into $\cW\cap\F_q^{\frac{q^r-1}{q-1}-r}$, 
incurring by \autoref{lem:redun} redundancy 
\[
r\parenv*{\cW\cap\F_q^{\frac{q^r-1}{q-1}-r}} 
\leq \parenv*{\frac{q^r-1}{q-1}-r}\frac{2}{k} \log_q\frac{q}{q-1}
.\]

Next, a systematic encoder for the $\sparenv*{\frac{q^r-1}{q-1},r,3}$ 
Hamming code (under the change of basis to $\bracenv*{\epsilon_i}$) 
can encode $\cW\cap\F_q^{\frac{q^r-1}{q-1}-r} \to 
\F_q^{\frac{q^r-1}{q-1}}$, incurring $r$ additional symbols of 
redundancy, and resulting in a code which can correct a single 
substitution.

Note, due to the systematic encoding, that the projection of this 
code onto the first $\frac{q^r-1}{q-1}-r$ coordinates is contained in 
$\cW$. We may simply cushion the last $r$ symbols with 
$\ceilenv*{\frac{2r}{k}}$ interleaved $1$'s (two per $k$ data symbols) 
to achieve a code $C\subseteq \cW\cap \F_q^n$ which may still correct 
a single substitution.
\end{IEEEproof}

Taking $n\to\infty$, we can compare the rate obtained by the code in 
\autoref{thm:ecc} to a simple upper bound of the best codes correcting 
only tandem duplications of length $k$ (see \cite{jain2017e}),
\[
R(C)\leq 1-\frac{(q-1)\log_q e}{q^{k+2}} + o(1).
\]
Clearly, then, a gap in rate exists, as 
$\frac{2}{k}\log_q\parenv*{\frac{q}{q-1}} > 
\frac{(q-1)}{q^{k+2}}\log_q(e) + o(1)$ for all $k\geq 2$. Note, 
however, that this upper bound is not necessarily tight, as it does 
not account for the combined error mode.

\section{Conclusion}\label{sec:conclusion}

We have studied the combination of a single substitution error with an 
unlimited number of tandem-duplication errors, with a fixed 
duplication-window length. We focused on two noise models, where the 
substitution error is either restricted to occur in an inserted copy 
during one of the duplication events, or may occur at any position in 
the string. We have presented bounds and a construction of 
error-detecting codes in the former error-model, as well as 
constructions of error-detecting and error-correcting codes in the 
latter.

In all cases, a rate loss is observed due to the need to recover from 
an unlimited number of duplications. Thus, we are interested in the 
\emph{extra redundancy cost} due to single-error detection or 
correction. In the first case, of detecting a single restricted 
substitution, we show that the additional required cost in redundancy 
is bounded from above by $\log_q(4(n-k))$ using a GV argument in 
\autoref{thm:1SDbounds}, where \autoref{Const:code} also shows that it 
is bounded from above by $\log_q(2(k+1)^2)$; depending on the 
asymptotic regime of $k$, either may be tighter than the other. In 
\autoref{Const:codeconstructive} we find a constructive procedure for 
generating codes for that purpose, which incur a higher redundancy 
cost of $4k\log_q(2)$; if $k$ is fixed, which is a likely scenario, 
then that cost is nonetheless constant as well, and improves upon 
\autoref{thm:1SDbounds}.

Further, in the second case of unrestricted substitution noise, 
\autoref{Const:1sdetcode} provides error-detecting codes for a single 
substitution incurring an extra redundancy cost of $O(\log(k^2 n))$. 
Finally, in the same error model, \autoref{cor:tan-sing_mut} and 
\autoref{thm:ecc} provide error-correcting codes which have lower 
rates than codes designed solely to correct duplication errors. 
Although we did not develop lower bounds on the required redundancy, 
it is our conjecture that both solutions offered here are sub-optimal. 
In particular, these latter codes rely on a constrained-coding 
approach which we do not believe is necessary in this context.
We also note that while both the upper bound and lower bound on the 
rates of these codes approach $1$ as $k\to\infty$, the lower bound 
does so as $\Theta(k^{-1})$ whereas the upper bound is much faster as 
$\Theta(q^{-k})$, implying a gap yet to be resolved. 

For future research, we would like to suggest a few generalizations of 
the noise model considered herein.
First, we suggest studying codes capable of handling a higher number 
of substitution errors. We also believe codes designed for handling 
only a bounded number of duplication events are of interest. 
Finally, we suggest to observe combinations of different noise 
mechanisms, including bounded tandem-duplication, end- or 
interspersed-duplication noise~\cite{hassanzadeh2016capacity}, or 
duplication and deletion noise.

\bibliographystyle{IEEEtranS}
%\bibliography{references}
%%%%%%%%%%%%%%%%%%%%%%%%%%%%%%%%%%%%%%%%%%%%%%%%%%%%%%%%%%%%%%%%%%%%%%
%%%%%%%%%%%%%%%%%%%%%%%%%%%%%%%%%%%%%%%%%%%%%%%%%%%%%%%%%%%%%%%%%%%%%%
% Generated by IEEEtranS.bst, version: 1.14 (2015/08/26)

%%%%%%%%%%%%%%%%%%%%%%%%%%%%%%%%%%%%%%%%%%%%%%%%%%%%%%%%%%%%%%%%%%%%%%
%%%%%%%%%%%%%%%%%%%%%%%%%%%%%%%%%%%%%%%%%%%%%%%%%%%%%%%%%%%%%%%%%%%%%%

\appendix

\begin{IEEEproof}[Proof of Lemma~\ref{lem:multi_ambi_subti_cases}]
\label{Proof:lemma_ambi_substitution}
Let $x'\in \dc^{*,\leq 1}(x)$, where 
\[
\abs{\rt(x')}=\abs{\rt(x)}, \qquad\text{but}\qquad \rt(x')\neq \rt(x),
\]
namely, an ambiguous unrestricted substitution occurred. Let us denote
\begin{align*}
    y &\eqdef \hat\phi (x), & z &\eqdef \bar\phi (x),\\
    y' &\eqdef \hat\phi (x'), & z' &\eqdef \bar\phi (x').
\end{align*}
Since duplications do not change the root, we assume without loss of 
generality that no duplications occur and only a single substitution 
occurs. Thus, we can write
\begin{align*}
    x'&=x+a\cdot e_i, & yz &= y'z'+a\cdot \epsilon_i,
\end{align*}
where $i$ denotes the location of the substitution, and 
$a\in{\Sigma\setminus\{0\}}$. Depending on $i$, a single substitution 
may result in one or two changed positions in the transform doamin of 
$\phi$. The proof of the claim comprises of many cases, and we start 
with some simple ones.

In the first simple case, the substitution occurs in the first $k$ 
positions, namely, $1\leq i\leq k$. Since $\phi(\rt(x'))=y'\mu(z')$, 
and $y\neq y'$, if we have $\abs{\rt(x')}=\abs{\rt(x)}$ then
\[ d(\phi(\rt(x)),\phi(\rt(x')))\leq 2,\]
by virtue of positions $i$ and $i+k$.

In a similar fashion, if the substitution occurs in the last $k$ 
positions, namely, $\abs{x}-k+1\leq i\leq \abs{x}$, only a single 
position is changed in the transform $\phi$. Since $\phi(\rt(x'))
=y'\mu(z')$, and $z\neq z'$, if we have $\abs{\rt(x')}=\abs{\rt(x)}$ 
then
\[ d(\phi(\rt(x)),\phi(\rt(x')))\leq 1,\]
by virtue of positions $i$.

We are now left with the last interesting case, in which the 
substitution changes two positions, $i$ and $i+k$, both in the $z$ 
part of the $\phi$-transform. We therefore disregard the part $y=y'$. 
We may now write
\[
\begingroup
\setlength\arraycolsep{2pt}
\begin{matrix}
z&=&u &a_1 &v &a_2 &w\\
z'&=& u& (a_1+a) &v& (a_2-a) & w
\end{matrix}
\endgroup
\]
where $u,w\in\Sigma^*$, $v\in\Sigma^{k-1}$, $a,a_1,a_2\in\Sigma$, and 
$a\neq 0$. We distinguish between two major cases, depending on 
whether $v= 0^{k-1}$.

\textbf{Case I:} In the first major case we have $v=0^{k-1}$. Let us 
write
\begin{align*}
    u&=u' 0^{m_1},& w&=0^{m_4}w',
\end{align*}
where all the indicated runs of zeros are maximal. Thus,
\[
\begingroup
\setlength\arraycolsep{2pt}
\begin{matrix}
z&=&u' &0^{m_1} &a_1 &0^{k-1} &a_2 &0^{m_4} &w'\\
z'&=&u' &0^{m_1} &(a_1+a) &0^{k-1} &(a_2-a) &0^{m_4} &w'.
\end{matrix}
\endgroup
\]
The length of the substring between $u'$ and $w'$ is $m_1+m_4+k+1$ and
we note that
\[
\floorenv*{\frac{m_1+m_4+k+1}{k}}=\floorenv*{\frac{m_1}{k}}
+\floorenv*{\frac{m_4}{k}}+s,
\]
where $s\in\{1,2\}$. We distinguish between the following cases:

    \begin{enumerate}
    \item If $a_1\neq 0$ and $a_2\neq 0$:
         \begin{enumerate}
             \item 
             If $a_1+a\neq 0$ and $a_2-a\neq 0$
                 \[ d(\phi(\rt(x)),\phi(\rt(x')))\leq 2.\]
             \item If exactly one of $a_1+a$ and $a_2-a$ is zero, the 
             length of $\mu(z')$ decreases by $k$.
             
             \item If $a_1+a=a_2-a=0$, the length of $\mu(z')$ 
             decreases by $sk$.
         \end{enumerate}
    \item If $a_1\neq 0$ and $a_2=0$:
         \begin{enumerate}
             \item If $a_1+a\neq 0$, since $a_2-a\neq 0$ the length of 
             $\mu(z')$ increases by $k$.
             
             \item If $a_1+a=0$, since $a_2-a\neq 0$
                 \[ d(\phi(\rt(x)),\phi(\rt(x')))\leq 1.\]
         \end{enumerate}
    \item If $a_1=0$ and $a_2\neq 0$:
         \begin{enumerate}
             \item If $a_2-a\neq 0$, since $a_1+a\neq 0$ the length of 
             $\mu(z')$ increases by $k$.
             
             \item If $a_2-a=0$, since $a_1+a\neq 0$
                 \[ d(\phi(\rt(x)),\phi(\rt(x')))\leq 1.\]
         \end{enumerate}
    \item If $a_1=a_2=0$, the length of $\mu(z')$ increases by $sk$.
\end{enumerate}

\textbf{Case II:} In the second major case, assume $v\neq 0^{k-1}$. 
Let us write
\[
u = u'0^{m_1}, \qquad v=0^{m_2}v' 0^{m_3}, \qquad w=0^{m_4}w',
\]
where all the indicated runs of zeros are maximal. Let 
$c\in\Sigma\setminus\{0\}$ be some nonzero letter in $v'$, important 
to us only for the purpose of being able to refer to the part of the 
string left of $c$ and the part of the string to the right of $c$.

\begin{enumerate}
    \item Examining the part of the string to the left of $c$:
          \begin{enumerate}
              \item If $a_1 \neq 0$: 
             \begin{enumerate}
                  \item If $a_1 = -a$:
                      \begin{enumerate}
                         \item \label{itm:ambi_a_dk} If 
                         $\floorenv*{\frac{m_{1}+m_{2}+1}{k}}
                         > \floorenv*{\frac{m_{1}}{k}}$, the length 
                         before $c$ decreases by $k$ and the substring 
                         $0^{j-1}(-a)0^{k-j}$ is deleted.
                         
                        \item \label{itm:ambi_a_01} If 
                        $\floorenv*{\frac{m_{1}+m_{2}+1}{k}}
                        = \floorenv*{\frac{m_{1}}{k}}$, the length 
                        before $c$ stays the same and the substitution 
                        $a_1\to 0$ occurs.
                     \end{enumerate}
                    \item \label{itm:ambi_a_02} If $a_1\neq -a$, the 
                    length before $c$ stays the same and the 
                    substitution $a_1\to (a_1+a)$ occurs.
             \end{enumerate}
              \item If $a_1 = 0$, then $a_1\neq -a$, and:
                    \begin{enumerate}
                        \item \label{itm:ambi_a_ik} If 
                        $\floorenv*{\frac{m_{1}+m_{2}+1}{k}}
                        > \floorenv*{\frac{m_{1}}{k}}$, the length 
                        before $c$ increases by $k$ and 
                        $0^{j-1}a0^{k-j}$ is inserted.
                        \item 
                        \label{itm:ambi_a_03} If 
                        $\floorenv*{\frac{m_{1}+m_{2}+1}{k}}
                        = \floorenv*{\frac{m_{1}}{k}}$, the length 
                        before $c$ stays the same and the substitution 
                        $0\to a$ occurs.
                    \end{enumerate}
          \end{enumerate}
    \item Examining the part of the string to the right of $c$:
         \begin{enumerate}
              \item If $a_2 \neq 0$:
                    \begin{enumerate}
                        \item If $a_2=a$:
                           \begin{enumerate}
                               \item \label{itm:ambi_b_dk}If 
                               $\floorenv*{\frac{m_{3}+m_{4}+1}{k}}
                               > \floorenv*{\frac{m_{4}}{k}}$, the 
                               length after $c$ decreases by $k$ and 
                               $0^{t-1}a0^{k-t}$ is deleted.
                               \item \label{itm:ambi_b_01}If 
                               $\floorenv*{\frac{m_{3}+m_{4}+1}{k}}
                               = \floorenv*{\frac{m_{4}}{k}}$, the 
                               length after $c$ stays the same and the 
                               substitution $a_2\to 0$ occurs.
                           \end{enumerate}
                        \item \label{itm:ambi_b_02} If $a_2\neq a$, 
                        the length after $c$ stays the same and the 
                        substitution $a_2\to (a_2-a)$ occurs.
                    \end{enumerate}
              \item If $a_2 = 0$, then $a_2\neq a$, and:
                    \begin{enumerate}
                         \item \label{itm:ambi_b_ik}If 
                         $\floorenv*{\frac{m_{3}+m_{4}+1}{k}}
                         > \floorenv*{\frac{m_{4}}{k}}$, the length 
                         after $c$ increases by $k$ and 
                         $0^{t-1}(-a)0^{k-t}$ is inserted.
                         \item \label{itm:ambi_b_03}If 
                         $\floorenv*{\frac{m_{3}+m_{4}+1}{k}}
                         = \floorenv*{\frac{m_{4}}{k}}$, the length 
                         after $c$ stays the same and the substitution 
                         $0\to (-a)$ occurs.
                    \end{enumerate}
          \end{enumerate}
\end{enumerate}
Based on the changes of $a_1$ and $a_2$, there are two types of 
ambiguous unrestricted substitutions:
\begin{itemize}
    \item
    Define the sets of cases $A\eqdef\{\text{\ref{itm:ambi_a_01}, 
    \ref{itm:ambi_a_02}, \ref{itm:ambi_a_03}}\}$ and 
    $B\eqdef\{\text{\ref{itm:ambi_b_01}, \ref{itm:ambi_b_02}, 
    \ref{itm:ambi_b_03}}\}$. Any substitution scenario from 
    $A\times B$ results in only two changed symbols, hence
    \[ d(\phi(\rt(x)),\phi(\rt(x')))\leq 2.\]
    \item
    The scenarios $(\text{\ref{itm:ambi_a_dk},\ref{itm:ambi_b_ik}})$ 
    and $(\text{\ref{itm:ambi_a_ik},\ref{itm:ambi_b_dk}})$ are more 
    complex because they involve both an inserted a substring and a 
    deleted  substring of length $k$. Since the two cases are similar, 
    we only show the analysis of the first case 
    $(\text{\ref{itm:ambi_a_dk},\ref{itm:ambi_b_ik}})$. We therefore 
    have
\[
\begingroup
\setlength\arraycolsep{2pt}
\begin{array}{ccccccccccc}
z&=&u' & 0^{m_1} & a &0^{m_2} &v' &0^{m_3} &0 &0^{m_4} &w'\\
z'&=&u' & 0^{m_1} & 0 &0^{m_2} &v' &0^{m_3} &a &0^{m_4} &w'
\end{array}
\endgroup
\]
    where we recall that $a\neq 0$, $\abs{v'}\leq k-1$, and $v'$ 
    starts and ends with a non-zero letter. Looking at $\mu(z')$ 
    compared with $\mu(z)$, the part to the left of $v'$ becomes 
    shorter by $k$ letters, whereas the part to the right of it 
    becomes longer by $k$ letters. In particular, we can write
\begin{equation}
\label{eq:swap}
\begingroup
\setlength\arraycolsep{2pt}
\begin{matrix}
\mu(z)&=&u'' &0^{\abs{v'}} &0^{m_3} a 0^{m_2} & v' &w''\\
\mu(z')&=& u''& v' &0^{m_3}a 0^{m_2} & 0^{\abs{v'}} & w''
\end{matrix}
\endgroup
\end{equation}
where $m_2+m_3+\abs{v'}+1=k$.
\end{itemize}

Having considered all cases, this last case is the only one in which 
we have an ambiguous unrestricted substitution in which potentially 
$d(\phi(\rt(x)),\phi(\rt(x')))\geq 3$. The swapping described 
in~\eqref{eq:swap} completes the proof of the claim.
\end{IEEEproof}

%%%%%%%%%%%%%%%%%%%%%%%%%%%%%%%%%%%%%%%%%%%%%%%%%%%%%%%%%%%%%%%%%%%%
%%%%%%%%%%%%%%%%%%%%%%%%%%%%%%%%%%%%%%%%%%%%%%%%%%%%%%%%%%%%%%%%%%%%
%%\begin{IEEEbiography}[{\includegraphics[width=1in,height=1.25in,clip,keepaspectratio]{mshell}}]{Michael Shell}
%% or if you just want to reserve a space for a photo:
%%\begin{IEEEbiography}{Michael Shell}
%%Biography text here.
%%\end{IEEEbiography}
%%
%% if you will not have a photo at all:
%\newpage
\begin{IEEEbiographynophoto}{Yuanyuan Tang}
is a Ph.D. candidate in the Department of Electrical and Computer 
Engineering at the University of Virginia. His research interests 
consist of information theory, coding theory, and wireless 
communications.

He received the Bachelor's degree in Engineering from the Department 
of Communication Engineering at Chongqing University in 2015 and the 
Master's degree in Engineering from the Department of Electronic 
Engineering at Tsinghua University in 2018.
\end{IEEEbiographynophoto}

\begin{IEEEbiographynophoto}{Yonatan Yehezkeally}
(S'12)
is a graduate student at the School of Electrical and Computer 
Engineering, Ben-Gurion University of the Negev, Beer-Sheva, Israel. 
His research interests include coding for DNA storage, combinatorial 
structures, algebraic coding and finite group theory.

Yonatan received the B.Sc.~(\emph{cum laude}) degree in Mathematics in 
2013, and the M.Sc.~(\emph{summa cum laude}) degree in Electrical and 
Computer Engineering in 2017, all from Ben-Gurion University of the 
Negev.
\end{IEEEbiographynophoto}

\begin{IEEEbiographynophoto}{Moshe Schwartz}
(M'03--SM'10)
is a professor at the School of Electrical and Computer
Engineering, Ben-Gurion University of the Negev, Israel. His research
interests include algebraic coding, combinatorial structures, and
digital sequences.

Prof.~Schwartz received the B.A.~(\emph{summa cum laude}), M.Sc., and
Ph.D.~degrees from the Technion -- Israel Institute of Technology,
Haifa, Israel, in 1997, 1998, and 2004 respectively, all from the
Computer Science Department. He was a Fulbright post-doctoral
researcher in the Department of Electrical and Computer Engineering,
University of California San Diego, and a post-doctoral researcher in
the Department of Electrical Engineering, California Institute of
Technology. While on sabbatical 2012--2014, he was a visiting 
scientist at the Massachusetts Institute of Technology (MIT).

Prof.~Schwartz received the 2009 IEEE Communications Society Best
Paper Award in Signal Processing and Coding for Data Storage, and the
2020 NVMW Persistant Impact Prize. He has also been serving as an
Associate Editor for Coding Techniques for the IEEE Transactions on
Information Theory since 2014.
\end{IEEEbiographynophoto}

\begin{IEEEbiographynophoto}{Farzad Farnoud (Hassanzadeh)}
(M'13)
is an Assistant Professor in the Department of Electrical and Computer 
Engineering and the Department of Computer Science at the University 
of Virginia. Previously, he was a postdoctoral scholar at the 
California Institute of Technology. 

He received his MS degree in Electrical and Computer Engineering from 
the University of Toronto in 2008. From the University of Illinois at 
Urbana-Champaign, he received his MS degree in mathematics and his 
Ph.D. in Electrical and Computer Engineering in 2012 and 2013, 
respectively. His current research interests include coding for data 
storage, data deduplication, and probabilistic modeling of genomic 
data for computational biology and data compression. He is the 
recipient of the 2013 Robert T. Chien Memorial Award from the 
University of Illinois for demonstrating excellence in research in 
electrical engineering and the recipient of the 2014 IEEE Data Storage 
Best Student Paper Award. 
\end{IEEEbiographynophoto}
%%%%%%%%%%%%%%%%%%%%%%%%%%%%%%%%%%%%%%%%%%%%%%%%%%%%%%%%%%%%%%%%%%%%
%%%%%%%%%%%%%%%%%%%%%%%%%%%%%%%%%%%%%%%%%%%%%%%%%%%%%%%%%%%%%%%%%%%%

\end{document}